\documentclass[lettersize,journal]{IEEEtran}
\usepackage{amsmath,amsfonts}
\usepackage{algorithmic}
\usepackage{algorithm}
\usepackage{array}
\usepackage[caption=false,font=footnotesize,labelfont=rm,textfont=rm]{subfig}
\usepackage{textcomp}
\usepackage{stfloats}
\usepackage{url}
\usepackage{verbatim}
\usepackage{graphicx}
\usepackage{cite}
\usepackage{comment}
\usepackage{ulem}
\usepackage{tabularray}
\usepackage{subcaption}
\usepackage{multirow}
\usepackage{xcolor}
\usepackage{tikz,xcolor}
\usepackage{orcidlink} 

\hypersetup{hidelinks}

\begin{document}

\title{Angle-Optimized Partial  Disentanglement for Multimodal Emotion Recognition in Conversation}

\author{Xinyi Che$^{\dagger}$$^{\orcidlink{0009-0008-7513-4501}}$, Wenbo Wang$^{\dagger}$$^{\orcidlink{0009-0004-7396-3956}}$, Yuanbo Hou$^{\orcidlink{0000-0001-8469-5740}}$, Mingjie Xie$^{\orcidlink{0000-0002-8399-6047}}$, Qijun Zhao$^{\orcidlink{0000-0003-4651-7163}}$, Jian Guan$^{\orcidlink{0000-0002-0945-1081}}$\IEEEmembership{Senior Member, IEEE}
\thanks{$^{\dagger}$Xinyi Che and Wenbo Wang contributed equally to this work}
\thanks{Corresponding author: Jian Guan}
\thanks{X. Che and Q. Zhao are with the School of Computer Science, Sichuan University, Chengdu 610065, China (e-mails: chexinyi@stu.scu.edu.cn; qjzhao@scu.edu.cn).}
  \thanks{W. Wang is with the Faculty of Computing, Harbin Institute of Technology, Harbin 150001, China (email: wwb1325864697@outlook.com).}
\thanks{Y. Hou is with the Machine Learning Group, Engineering Science, University of Oxford, U.K. (e-mail: yuanbo.hou@eng.ox.ac.uk).}
\thanks{M. Xie is with the School of Astronautics, Beihang University, Beijing 100191, China (e-mail: xiemingjie@buaa.edu.cn).}
  \thanks{J. Guan is with the Group of Intelligent Signal Processing (GISP), College of Computer Science and Technology, Harbin Engineering University, Harbin 150001, China (e-mail: j.guan@hrbeu.edu.cn).}
}

\markboth{Journal of \LaTeX\ Class Files,~Vol.~14, No.~8, August~2021}
{Shell \MakeLowercase{\textit{et al.}}: A Sample Article Using IEEEtran.cls for IEEE Journals}

\maketitle

\begin{abstract}
Multimodal Emotion Recognition in Conversation (MERC) aims to enhance emotion understanding by integrating complementary cues from text, audio, and visual modalities. Existing MERC approaches predominantly focus on cross-modal shared features, often overlooking modality-specific features that capture subtle yet critical emotional cues such as micro-expressions, prosodic variations, and sarcasm. Although related work in multimodal emotion recognition (MER) has explored disentangling shared and modality-specific features, these methods typically employ rigid orthogonal constraints to achieve full disentanglement, which neglects the inherent complementarity between feature types and may limit recognition performance. 
To address these challenges, we propose Angle-Optimized Feature Learning (AO-FL), a framework tailored for MERC that achieves partial disentanglement of shared and specific features within each modality through adaptive angular optimization. Specifically, AO-FL aligns shared features across modalities to ensure semantic consistency, and within each modality it adaptively models the angular relationship between its shared and modality-specific features to preserve both distinctiveness and complementarity. An orthogonal projection refinement further removes redundancy in specific features and enriches shared features with contextual information, yielding more discriminative multimodal representations. Extensive experiments confirm the effectiveness of AO-FL for MERC, demonstrating superior performance over state-of-the-art approaches. Moreover, AO-FL can be seamlessly integrated with various unimodal feature extractors and extended to other multimodal fusion tasks, such as MER, thereby highlighting its strong generalization beyond MERC.

\end{abstract}

\begin{IEEEkeywords}
adaptive angle optimization, feature partial disentanglement, orthogonal projection refinement, multimodal emotion recognition in conversation.
\end{IEEEkeywords}
\section{Introduction}
\begin{figure}[t!]
    \centering
    \includegraphics[width=0.95\linewidth]{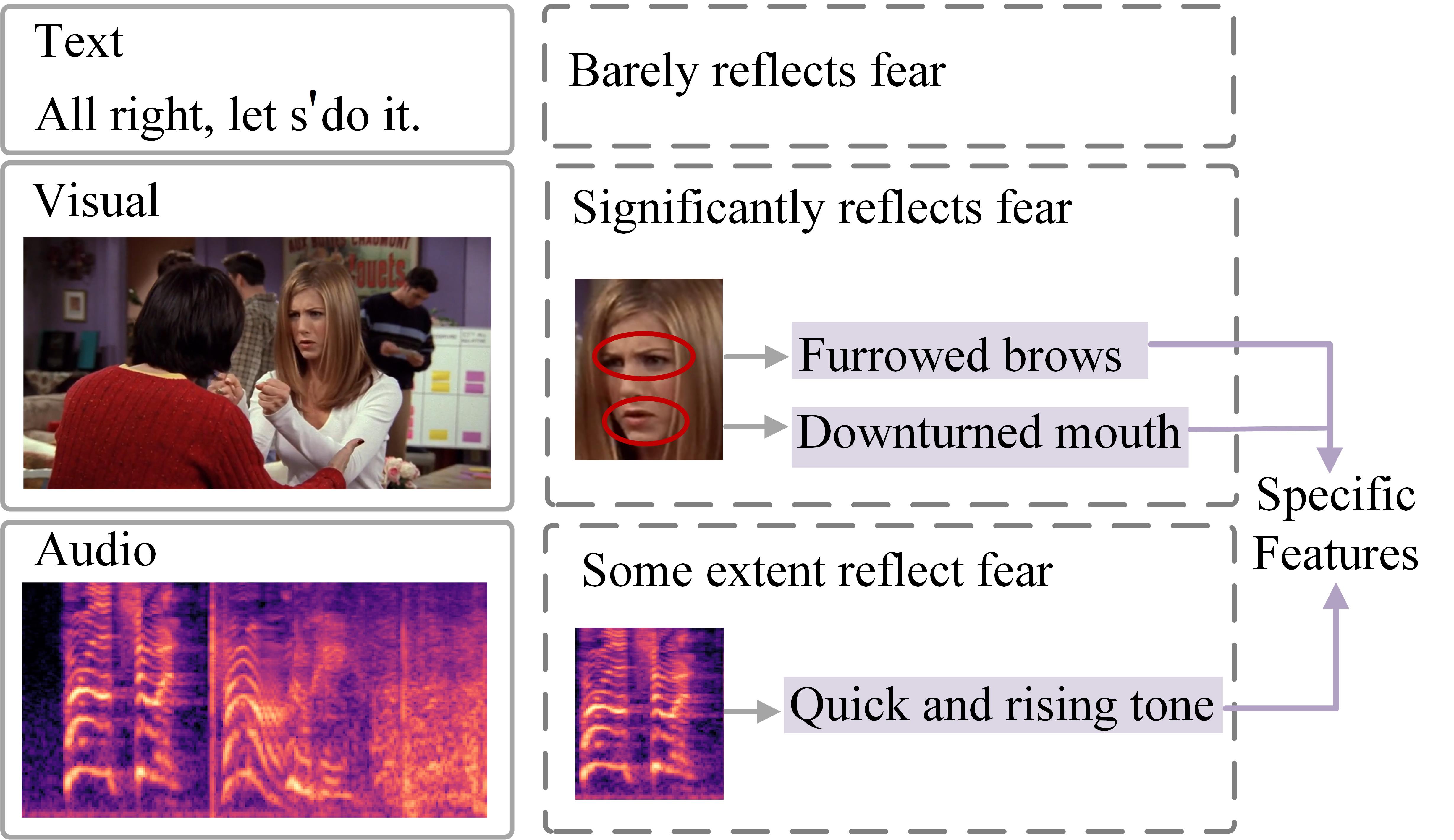}
    \caption{An example from the MELD \cite{poria-etal-2019-meld} dataset to illustrate specific features. For instance, the semantics of fear can be reflected in specific features, such as furrowed brows and a downturned mouth in the visual modality, or a quick and rising tone in the audio modality.}
    \label{figure1}
\end{figure}
\IEEEPARstart{M}{ultimodal} Emotion Recognition in Conversation (MERC) aims to enhance emotion recognition by effectively integrating complementary information across textual, audio, and visual modalities \cite{jing2024dq,wang2024dynamic}. Compared to traditional unimodal emotion recognition methods, MERC achieves significantly better performance due to its capability to capture rich and diverse emotional cues from different modalities \cite{ghosal-etal-2019-dialoguegcn, hu-etal-2021-mmgcn,fu2025himul}. This can be beneficial for applications in human-computer interaction, affective computing, where accurate emotion recognition enhances user experience and enables more intuitive communication \cite{DBLP:conf/acl/ZhaoLK20, li-etal-2022-enhancing-knowledge}.

Recent approaches \cite{ghosal-etal-2019-dialoguegcn, wei2019mmgcn, hu2022mm, li-etal-2023-joyful, li2023decoupled, meng2024masked} to MERC often focus on capturing shared features that are consistent across different modalities. For example, emotional sentiments reflected through spoken words, facial expressions, and voice intonation can often be semantically aligned. However, such methods often overlook modality-specific features, which contain modality-specific semantic cues, such as nuanced tonal variations, subtle facial micro-expressions, or sarcasm conveyed through prosody. Ignoring these specific features limits the model's ability to fully capture the richness of emotional cues, thereby affecting the MERC performance.

Many existing methods \cite{wei2019mmgcn, hu2022mm, ghosal-etal-2019-dialoguegcn, guo2024speaker, lian2021ctnet, meng2024masked, majumder2019dialoguernn, yang2023self} simply aggregate features from different modalities based on semantic similarity. For instance, MMGCN \cite{wei2019mmgcn} and MMDFN \cite{hu2022mm} treat features from all modalities as nodes, using semantic similarity as weighted edges to aggregate emotion-related features via graph neural networks (GNNs) \cite{morris2019weisfeiler}. Aggregating modality features purely by semantic alignment, without explicitly distinguishing shared and modality-specific components within each modality, will introduce redundancy and hinder performance. Joyful \cite{li-etal-2023-joyful} improves recognition by emphasizing shared semantics, projecting unimodal features into a common subspace while treating modality-specific information as noise. However, this overemphasis on shared features overlooks important modality-specific cues for the MERC task, limiting performance gains.
This is because, as illustrated in Fig.~\ref{figure1}, specific semantics, such as \textit{furrowed brows} and a \textit{downturned mouth} in video, combined with a \textit{quick, rising tone} in audio, can also contribute to the recognition of emotions like \textit{fear}.

Some methods from the broader multimodal emotion recognition (MER) task \cite{hazarika2020misa, shi2024co, li2023decoupled, wang2025dlf, chen2025dual, yang2023confede, yang2022disentangled} attempt to simultaneously model shared and modality-specific features. Approaches such as MISA \cite{hazarika2020misa} and DMD \cite{li2023decoupled} achieve this via strict orthogonality (i.e., $90^\circ$) constraints to fully disentangle the two feature types, treating them as entirely distinct patterns. While this enforces distinctiveness, it neglects their complementarity, as both features may convey aspects of the same underlying emotion, thereby limiting performance.

To address the above limitations, we propose an Angle-Optimized Feature Learning (AO-FL) framework that performs partial disentanglement of shared and modality-specific features within each modality, ensuring both distinctiveness and complementarity for the MERC task. 
Our AO-FL consists of an Adaptive Angle Optimization (AAO) strategy and an Orthogonal Projection Refinement (OPR) module. The AAO strategy learns adaptive angular relationships ($0^\circ$–$180^\circ$) between shared and specific features of the same modality, allowing them to be sufficiently separated to mitigate redundancy while remaining correlated enough to retain complementary emotion cues.
Within AAO, a contrastive-based consistency enhancement function first aligns shared features across modalities to ensure semantic consistency, followed by an angular relationship exploration module that models these angular relationships in a self-supervised manner to achieve partial disentanglement.
Meanwhile, the OPR module further enhances the partial disentangled features by projecting modality-specific features to remove redundancy, 
thereby improving their compatibility with shared features for more effective fusion. In addition, it enriches shared features with contextual information through a self-attention mechanism, enhancing their representational capacity.
Extensive experiments conducted on IEMOCAP \cite{busso2008iemocap} and MELD \cite{poria-etal-2019-meld} datasets validate the effectiveness and superiority of our AO-FL for the MERC task, while ablation studies verify the contribution of each component and the benefits of its angle-optimized partial disentanglement strategy.
Furthermore, our AO-FL can be seamlessly integrated with advanced LLM-based unimodal feature extractors \cite{lian2025affectgpt, ge2024early, fan2024leveraging} and applied to broader MER task, highlighting its strong generalization.

The contributions of our study are summarized as follows:
\begin{itemize}
    \item {Proposing the AO-FL framework, which integrates an AAO strategy with an OPR module to capture complementary shared and specific features for MERC via partial disentanglement.}
    \item { Developing an AAO strategy to learn adaptive angular relationships between shared and specific features within each modality, enabling sufficient separation to reduce redundancy while retaining complementary emotion cues.}
    \item Designing an OPR module to reduce redundancy of specific feature, and enhance semantics of shared features, thereby significantly improving recognition accuracy. 
    \item Achieving state-of-the-art performance on MERC benchmarks (IEMOCAP and MELD), validating the effectiveness of AO-FL for MERC, and extending to broader MER tasks by seamlessly integrating with advanced unimodal feature extractors, demonstrating strong generalization.
\end{itemize}

The remainder of the paper is organized as follows: Section~\ref{sec:related} introduces the related work of our proposed method; Section~\ref{sec:method} presents the proposed method in detail; Section~\ref{sec:experiment} shows the experimental results; and Section~\ref{sec:conclusion} summarizes the paper and gives the conclusion.

\section{Related Work}
\label{sec:related}
In this section, we categorize existing MERC methods into non-disentangled and disentangled approaches, based on whether or not they explicitly distinguish cross-modal shared features and modality-specific features for multimodal fusion.

\subsection{Non-Disentangled Methods}

Non-disentangled methods for MERC mainly focus on directly aggregating emotional features across modalities, typically relying on semantic similarity without explicitly separating shared and modality-specific semantics~\cite{wei2019mmgcn, hu2022mm, ghosal-etal-2019-dialoguegcn, guo2024speaker, lian2021ctnet, meng2024masked, majumder2019dialoguernn, yang2023self,shi2023multiemo,meng2024revisiting, liu2025tacfn, li2023revisiting,shou2024adversarial,lu2024hypergraph,zhang2024rl}.

For instance, DialogRNN \cite{majumder2019dialoguernn} leverages Gated Recurrent Units (GRUs) to model inter-utterance contextual dependencies, integrating multimodal features without explicitly distinguishing their modality-specific contributions.
DialogGCN \cite{ghosal-etal-2019-dialoguegcn} builds a graph from sequentially encoded utterances, with multimodal features as nodes and utterance relationships as edges, enhancing speaker-level context modeling.
Similarly, MMGCN \cite{wei2019mmgcn} and MMDFN \cite{hu2022mm} formulate emotion recognition as graph-based problems, aggregating features using graph convolution without explicitly separating implicit modality-specific semantics. 
GS-MCC \cite{meng2024revisiting} builds multimodal semantic interaction graphs and employs Fourier-based Graph Neural Networks to capture long-range dependencies and frequency-domain semantic correlations.
Moreover, Transformer-based methods such as CTNet \cite{lian2021ctnet}, SCMM \cite{yang2023self} and SACCMA \cite{guo2024speaker} fuse features across modalities through cross-modal attention mechanisms to leverage context effectively. MGLRA \cite{meng2024masked} integrates attention mechanisms combined with GCN for intra-modal and cross-modal feature fusion, enhancing multimodal representation learning for MERC. Additionally, some methods incorporate external auxiliary information, such as pre-trained language knowledge \cite{hu2024unimeec}, \cite{tu2024multiple} and personality information \cite{tu2024persona}, to further improve emotion recognition performance.

However, all these non-disentangled methods aggregate multimodal information solely based on semantic similarities without explicitly distinguishing shared and implicit modality-specific semantics, leading to redundancy and limited recognition performance.
\begin{figure*}[t]
    \centering
    \includegraphics[width=.98\linewidth]{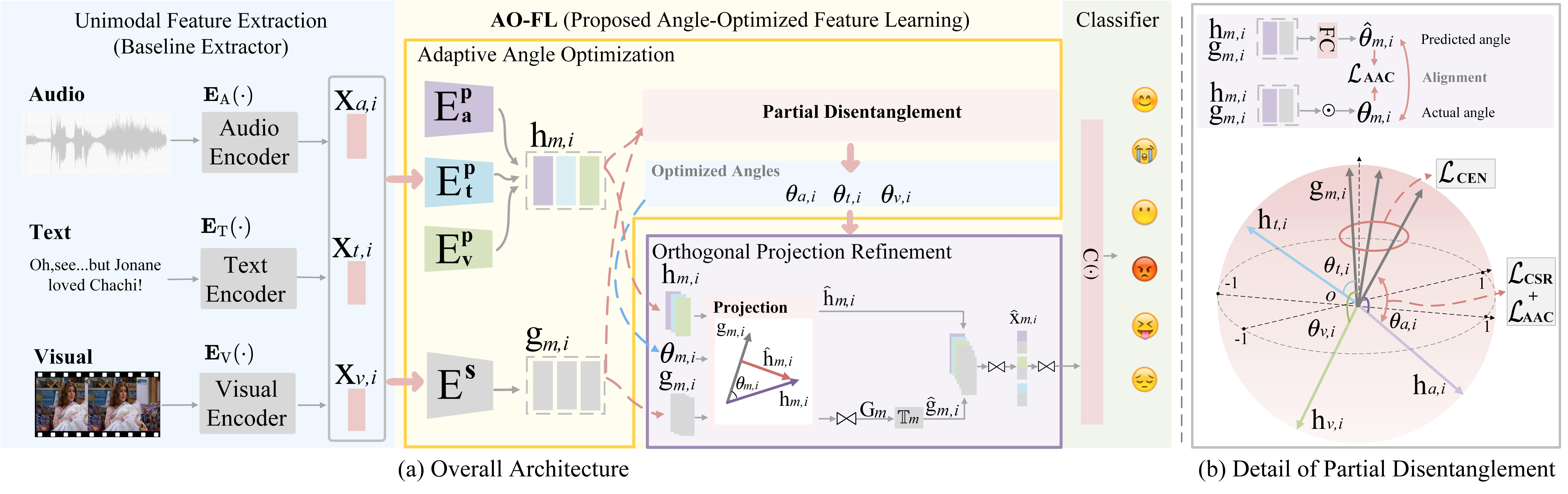}
    \caption{Overview of the proposed Angle-Optimized Feature Learning (AO-FL) framework for MERC.
    (a) Overall Architecture: a baseline unimodal feature extractor (MultiEMO \cite{shi2023multiemo}), followed by AO-FL, which comprises Adaptive Angle Optimization (AAO) and Orthogonal Projection Refinement (OPR) for partial disentanglement, and the final classifier. 
    The subscript $m$ ($m \in {a, t, v}$) denotes the audio, text, or visual modality. $\bowtie$ denotes concatenation.
    (b) Partial Disentanglement detail: Shared and specific features from each modality are adjusted to optimize angles ($\theta_{a,i}$,$\theta_{t,i}$,$\theta_{v,i}$) within  $0 ^\circ$ to $180^\circ$. Three losses jointly guide angle optimization and feature separation, where $\mathcal{L}_{CEN}$ aligns shared features across modalities, $ \mathcal{L}_{AAC}$ adaptively optimizes the angle between shared and specific features, and $\mathcal{L}_{CSR}$ ensures the learned angle is sufficiently large to distinguish shared and specific features.
    $\odot$ denotes cosine similarity calculation.
    }
    \label{multimodal_2}
\end{figure*}[t]

\subsection{Disentangled Methods}

Disentangled methods aim to decompose multimodal features into two complementary components: a shared feature that captures cross-modal consistent semantics, and a modality-specific feature that preserves information unique to each modality~\cite{li-etal-2023-joyful, hazarika2020misa, shi2024co, li2023decoupled, wang2025dlf, chen2025dual, yang2023confede, yang2022disentangled}.  

For the MERC task, although Joyful \cite{li-etal-2023-joyful} distinguishes shared and specific features within each modality, it projects unimodal features from all modalities into a unified space to obtain shared features across modalities, while treating modality-specific information as noise. This design neglects unique modality-specific cues, thereby limiting recognition performance.  Other studies \cite{dai2024multimodal, su2024dynamic, li2023revisiting} have attempted to disentangle unimodal features from different perspectives, such as separating them into emotion-relevant and emotion-irrelevant components \cite{su2024dynamic}. However, these works primarily focus on reducing redundancy across modalities to improve fusion, without performing fine-grained, modality-level analysis to identify the shared and modality-specific patterns of each unimodal feature for the MERC task.  

In contrast, several related works in MER \cite{hazarika2020misa, shi2024co, li2023decoupled, wang2025dlf, chen2025dual, yang2023confede, yang2022disentangled} explicitly perform decomposition of shared and modality-specific features. These approaches typically separate each modality’s unimodal feature into shared and specific components via full disentanglement constraints ($90^\circ$) to capture different patterns of a specific modality, and then jointly consider both feature types for emotion recognition. For instance, MISA \cite{hazarika2020misa} and CRNet \cite{shi2024co} employ the squared Frobenius norm to enforce orthogonality between shared and specific features of the same modality, whereas DMD \cite{li2023decoupled}, DLF \cite{wang2025dlf}, and DSCN \cite{chen2025dual} minimize their cosine similarity to achieve the same effect. Nevertheless, all these orthogonality-based full disentanglement methods focus primarily on strengthening feature distinctiveness, while overlooking their complementarity. Overemphasizing distinctiveness risks discarding emotion-related information, as both feature types inherently reflect the same underlying emotion.

Different from previous works
~\cite{li-etal-2023-joyful, dai2024multimodal,
hazarika2020misa, shi2024co, li2023decoupled, wang2025dlf, chen2025dual, yang2023confede, yang2022disentangled},  
our proposed AO-FL obtains partially disentangled shared and modality-specific features within each modality, and adaptively models their angular relationship to simultaneously preserve distinctiveness and complementarity. This design not only addresses the limitations of full disentanglement for the MERC task, but also demonstrates strong generalization ability across different unimodal feature extractors and other multimodal fusion tasks, as shown in Section~\ref{sec:MER} of our experiments.

\section{Proposed Method}
\label{sec:method}

In this section, we present the details of unimodal feature extraction, our Angle-Optimized Feature Learning (AO-FL) and emotion recognition. Specifically, the two components of our proposed AO-FL, i.e., Adaptive Angle Optimization (AAO) and Orthogonal Projection Refinement (OPR), are introduced in detail. 
The overview is illustrated in Fig.~\ref{multimodal_2}.
\subsection{Unimodal Feature Extraction}
Given a conversation set $\mho = \{\mathcal{U}_{1},\dots, \mathcal{U}_{n},\dots ,\mathcal{U}_{N}\}$ consisting of $N$ utterances, each utterance $\mathcal{U}_{n}$ is represented as a tuple $\mathcal{U}_{n} = (\mathcal{U}_{a, n}, \mathcal{U}_{t, n}, \mathcal{U}_{v, n})$, corresponding to audio, text and visual modalities, respectively.

Following the MultiEMO \cite{shi2023multiemo}, we also employ three pre-trained encoders, denoted as $\mathbf{E}_\text{A}(\cdot)$, $\mathbf{E}_\text{T}(\cdot)$, and $\mathbf{E}_\text{V}(\cdot)$, to extract unimodal features (i.e., $\mathbf{X}_{a}$, $\mathbf{X}_{t}$, and $\mathbf{X}_{v} \in \mathbb{R}^{N \times d}$) from the audio, text, and visual modalities at conversation-level (i.e., $\mho_a , \mho_t$ and $\mho_v$), respectively, as follows,
\begin{equation}
\label{eq:3}
\begin{split}
\mho_a &= \{\mathcal{U}_{a, 1},\dots, \mathcal{U}_{a, n},\dots ,\mathcal{U}_{a, N}\},\\ 
\mho_t &= \{\mathcal{U}_{t, 1},\dots, \mathcal{U}_{t, n},\dots ,\mathcal{U}_{t, N}\},\\
\mho_v &= \{\mathcal{U}_{v, 1},\dots, \mathcal{U}_{v, n},\dots ,\mathcal{U}_{v, N}\}
\end{split}
\end{equation}
\begin{equation}
    \mathbf{X}_{a} = \mathbf{E}_\text{A}(\mho_a), \quad
    \mathbf{X}_{t} = \mathbf{E}_\text{T}(\mho_t), \quad
    \mathbf{X}_{v} = \mathbf{E}_\text{V}(\mho_v)
\end{equation}
where $d$ is the dimension of the unimodal features.

Specifically,  $\mathbf{E}_\text{A}(\cdot)$ consists of the OpenSMILE toolkit \cite{eyben2010opensmile} for low-level audio feature extraction and DialogRNN \cite{majumder2019dialoguernn} for modeling contextual dependencies across utterances. 
The text encoder $\mathbf{E}_\text{T}(\cdot)$ is based on a RoBERTa \cite{kim2021emoberta}, which generates context-independent feature vectors for each utterance, followed by a fully connected layer to obtain utterance-level representation. For the visual modality, the encoder $\mathbf{E}_\text{V}(\cdot)$
is a joint model composed of VisExtNet \cite{shi2023multiemo} for visual feature extraction and DialogRNN for temporal modeling.

To facilitate the descriptions in the following sections, we leverage $\textbf{x}_{m,i}\in \mathbb{R}^d$ to denote the unimodal feature of utterance $i$ in conversation $\mho$, where $m \in\left \{a, t, v\right \}$ represents the modality, and $i$ is the utterance index.

Notably, we only take the extractors of MultiEMO \cite{shi2023multiemo} as an example here. Various unimodal feature extractors, including advanced LLM-based feature extractors \cite{yang2023baichuan, hsu2021hubert, radford2021learning}, can be compatible with our AO-FL to solve different tasks (i.e., MER), as shown in Section \ref{sec:MER}.

\subsection{Adaptive Angle Optimization}
To ensure cross-modal semantic consistency while preserving modality-specific distinctions, we propose an Adaptive Angle Optimization (AAO) strategy. It first extracts shared and specific features from the unimodal feature of each modality, then aligns shared features across modalities, and finally models the angular relationship between shared and specific features, achieving partial disentanglement.
\subsubsection{Shared and Modality-Specific Feature Extraction} 
To capture the consistent semantics across modalities, a shared encoder \(\mathbf{E}^{\text{s}}(\cdot)\) is used to extract shared features for all modalities:
\begin{equation}
    \mathbf{g}_{m,i} = \mathbf{E}^{\text{s}}(\mathbf{x}_{m,i}; \lambda)
\end{equation}
where \(\mathbf{g}_{m,i} \in \mathbb{R}^d\) denotes the shared feature of the \(i\)-th utterance in modality \(m\), and \(\lambda\) represents the parameters of the shared encoder, which consists of two fully connected layers.

To preserve the modality-specific semantics, we employ three modality-specific  encoders $\mathbf{E}^{\text{p}}_{m}$ to extract specific features individually for each modality:
\begin{equation}
    \mathbf{h}_{m,i} = \mathbf{E}^{\text{p}}_{m}(\mathbf{x}_{m,i}; \psi_{m})
\end{equation}
where \(\mathbf{h}_{m,i} \in \mathbb{R}^d\) denotes the specific feature of modality \(m\), and \(\psi_m\) refers to the parameters of the corresponding modality-specific encoder,  which is also realized via two fully connected layers.
\begin{figure}
\centering
 \includegraphics[width=0.75\linewidth]{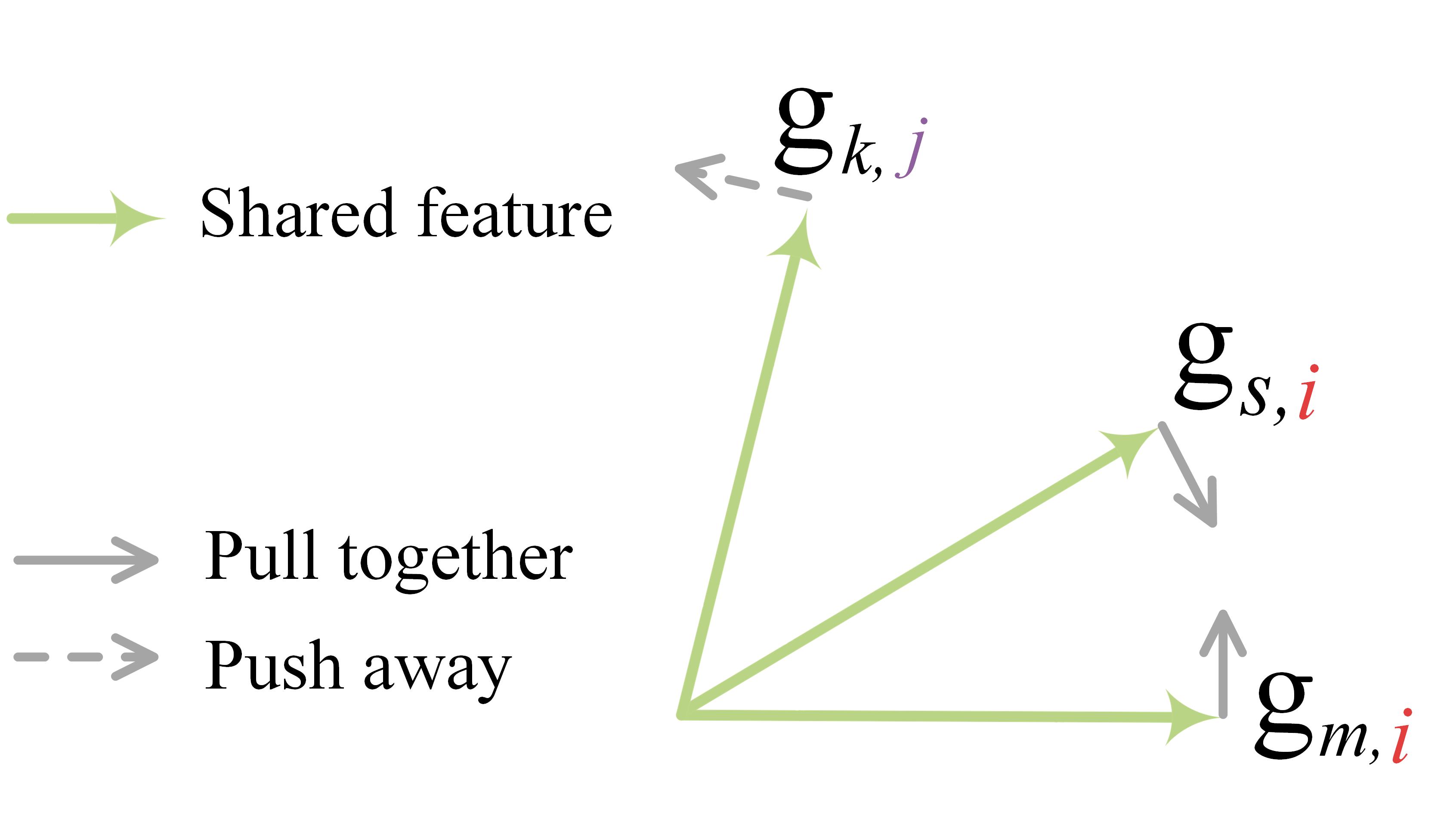}
\caption{Illustration of cross-model consistency enhancement, where shared features (i.e., $\mathbf{g}_{s, i}$ and $\mathbf{g}_{m, i}$) from different modalities ($s \neq m$) within the same utterance  are pulled together as positive pairs, while those from different utterances ($i \neq j$) with mismatched emotion labels (i.e., $\mathbf{g}_{k, j}$) are pushed apart as negative pairs. ($\{s, m$ and $k\in \{a, t, v\}\}$)}
\label{CE_loss}
\end{figure}
\subsubsection{Cross-Modal Consistency Enhancement}
To avoid semantic inconsistency across modalities, we employ a 
Consistency Enhancement (CEN) function based on contrastive learning, explicitly aligning shared features across modalities.

Specifically, for the $i$-th utterance, shared features from different modalities of the same utterance are treated as positive pairs, while shared features from different utterances with mismatched emotion labels form negative pairs. The contrastive consistency enhancement loss $\mathcal{L}_{CEN} $ is defined as:
\begin{equation}
\label{eq:constrastive}
\small
\mathcal{L}_{CEN} = -\log
\frac{\exp{({\mathbf{g}_{m,i}}^{\top} \cdot \mathbf{g}_{s,i})}}{\exp({\mathbf{g}_{m,i}}^{\top} \cdot \mathbf{g}_{s,i})
 + \sum_{\mathbf{g}_{k,j} \in \Omega_{m,i}} \exp({\mathbf{g}_{m,i}}^{\top} \cdot \mathbf{g}_{k,j})}
\end{equation}
where $s \neq m$, $i \neq j$, $\{s, m$ and $k\in \{a, t, v\}\}$. ${\top}$ denotes transpose operation, and  $\Omega_{m,i}$ represents the negative sample set for shared feature $\mathbf{g}_{m, i}$. 
Minimizing $\mathcal{L}_{CEN}$ enhances cross-modal consistency by pulling the shared features of corresponding utterances closer while pushing mismatched pairs further apart, as illustrated in Fig.~\ref{CE_loss}.  
Consequently, we can obtain the shared features for the same utterance, containing highly consistent semantics across different modalities.

\subsubsection{Within-Modality Angle Exploration} 
To capture the relationship between the shared and specific features of the same modality with an adaptive angle, the Angular Relationship Exploration (ARE) module introduces an Adaptive Angle Constraint (AAC) and a Cosine Similarity Ranking (CSR) criterion, which can preserve complementary emotion-related semantics between them while enhancing their distinctions.

\noindent\textbf{Adaptive Angle Constraint}: 
We first measure the actual angular relationship between shared feature $\mathbf{g}_{m,i}$ and the specific feature $\mathbf{h}_{m,i}$, as follows:
\begin{equation}
    \cos(\theta_{m,i}) = \frac{\mathbf{g}_{m,i}\cdot{}\mathbf{h}_{m,i} }{|| \mathbf{g}_{m,i}|| \cdot ||\mathbf{h}_{m,i}||} 
\end{equation}
where $\theta_{m,i}$ is the actual angle of these two features.

A learnable angle  $\hat{\theta}_{m,i}$ is then predicted  with a fully connected layer: 
\begin{equation}
 \cos(\hat{\theta}_{m,i}) = \tanh (\mathbf{W} \cdot [\mathbf{g}_{m,i},\mathbf{h}_{m,i}]^{\top}+\mathbf{b})
\end{equation}
where $\mathbf{W} \in \mathbb{R}^{1\times 2d}$ and $\mathbf{b} \in \mathbb{R}^{1\times 1}$ are the linear mapping matrix and bias, respectively. $[\cdot, \cdot]$ denotes the concatenation operation.

An AAC loss $\mathcal{L}_{AAC}$ is utilized to align actual and predicted angles of shared and specific features in a self-supervised manner using root mean squared error (RMSE):
\begin{equation}
  \mathcal{L}_{AAC} = \text{RMSE}(\cos (\theta_{m, i}), \cos( \hat{\theta}_{m,i})) 
\label{Correction}
\end{equation}

Unlike the rigid orthogonality constraint (i.e., $90^\circ$)  between shared and specific features \cite{hazarika2020misa}, our AAC loss adaptively adjusts their angular relationship within $0^\circ$–$180^\circ$, enabling gradual optimization. This flexibility better captures both cross-modal consistent semantics and modality-specific distinctions.

\begin{figure}
\centering
 \includegraphics[width=0.75\linewidth]{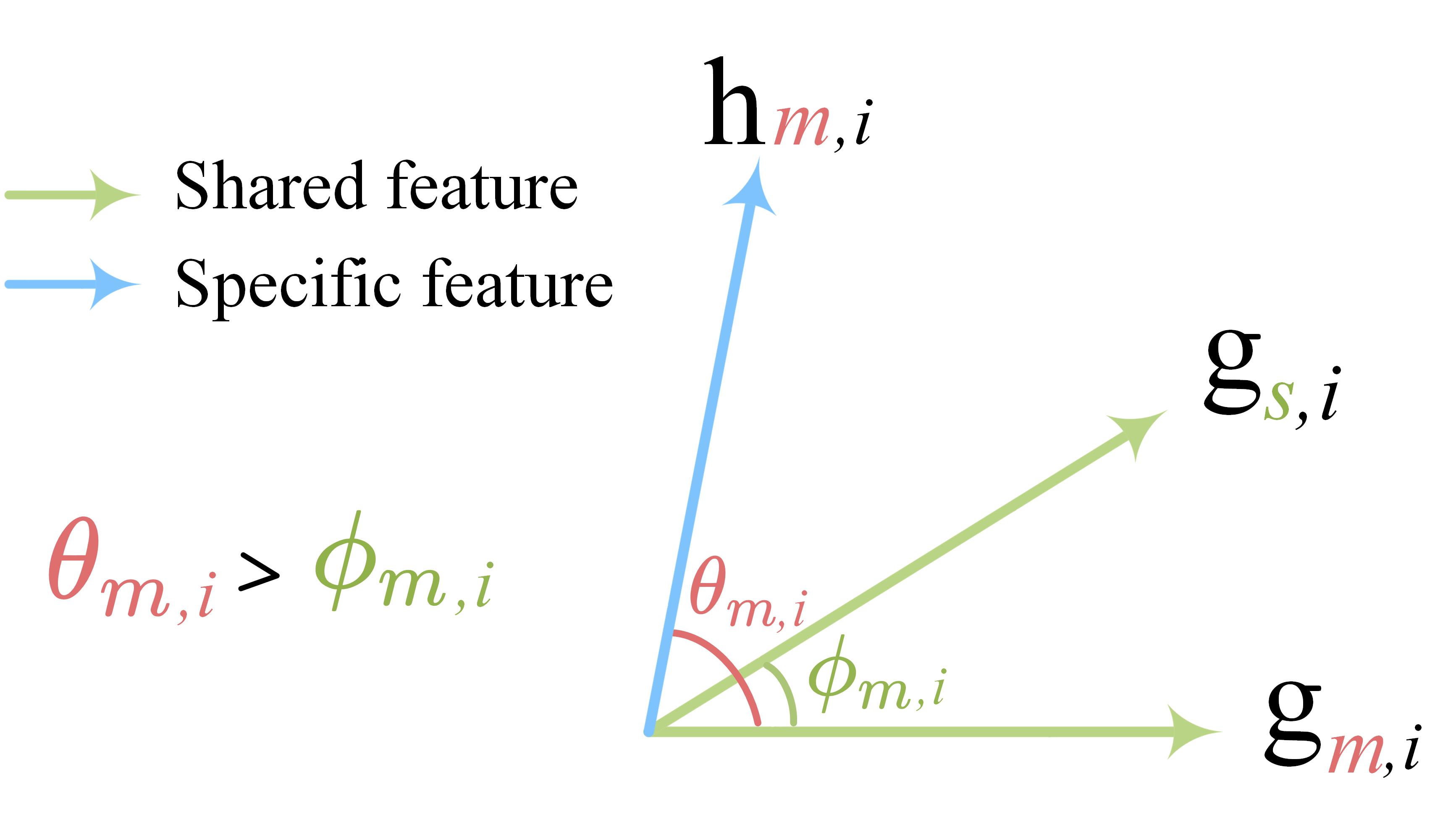}
 \caption{Illustration of the cosine similarity ranking criterion, where $\mathbf{h}_{m,i}$ denotes the specific feature and $\mathbf{g}_{s,i}$, $\mathbf{g}_{m,i}$ are the shared features. 
 The criterion enforces that the angle between the same utterance's shared features from different modalities ($\phi_{m,i}$) is smaller than that between the specific and shared features of a specific modalities ($\theta_{m,i}$), i.e., $\theta_{m,i} > \phi_{m,i}$.
 }
    \label{fig:CosRank}
\label{CSR_loss}
\end{figure}

\noindent\textbf{Cosine Similarity Ranking:}  
To prevent the learned angle from becoming too small to distinguish shared and specific features, we introduce a cosine similarity ranking criterion. 

Concretely, for each utterance $i$, we enforce that the angle between shared and specific features of the same modality remains larger than that among shared features across different modalities, as illustrated in Fig. \ref{CSR_loss}. Formally, a CSR loss $\mathcal{L}_{CSR}$ is defined as:
\begin{equation}
  \mathcal{L}_{CSR} = \max(\cos({\phi}_{m,i}) -\cos({\theta}_{m,i}), 0) 
\label{Correction}
\end{equation}
where $\cos({\theta}_{m,i})$ measures the similarity between $\mathbf{g}_{m,i}$ and $\mathbf{h}_{m,i}$ of the same modality, while $\cos({\phi}_{m,i})$ indicates the similarity among shared features from different modalities as: 
\begin{equation}
    \cos(\phi_{m,i}) = \frac{\mathbf{g}_{m,i}\cdot{}\mathbf{g}_{s,i} }{\lvert |  \mathbf{g}_{m,i}\rvert | \cdot  \lvert  | \mathbf{g}_{s,i}  \rvert |  } 
\end{equation}
where $\mathbf{g}_{m,i}$ and $\mathbf{g}_{s,i}$ denote the $i$-th utterance's shared features from modalities $m$ and $s$ ($m \neq s$), respectively. 
Minimizing this loss ensures that shared and specific features are partially disentanglement to capture distinct features.

The overall ARE loss is a weighted combination of the AAC and CSR, expressed as:
\begin{equation}
   \mathcal{L}_{ARE} = \gamma \cdot \mathcal{L}_{AAC} + \mu \cdot \mathcal{L}_{CSR}
\end{equation}
where $\gamma$ and $\mu$ are the penalty parameters. 

By optimizing $\mathcal{L}_{ARE}$, we achieve partial disentanglement, effectively balancing cross-modal consistent semantics and modality-specific unique semantics,
as case study demonstrated in Section~\ref{sec:case}.

\subsection{Orthogonal Projection Refinement}
To further enhance the cross-modal consistent semantics and modality-specific distinct semantics, we propose an Orthogonal Projection Refinement (OPR) module to reduce the redundancy between the shared and specific features via a projection operation, as follows:
\begin{equation}
\mathbf{\hat{h}}_{m,i}  = {\mathbf{h}_{m,i}} -{\mathbf{h}_{m,i}\cdot \cos (\theta_{m,i})}
\end{equation}
Thus, we can obtain a refined specific feature $\mathbf{\hat{h}}_{m,i}$ that only retains the distinct specific-modality semantics by subtracting its redundancy part according to the angular relationship $\theta_{m,i}$.

Meanwhile, to enhance the partially disentangled shared features with contextual semantics in modality $m$, an $L$-layer Transformer (i.e., $\mathbf{T}_m(\cdot)$) with self-attention mechanism \cite{vaswani2017attention} is applied:
\begin{equation}
\mathbf{G}_m = [\mathbf{g}_{m,0}, \cdots, \mathbf{g}_{m,i}, \cdots \mathbf{g}_{m,N}] \in \mathbb{R}^{N\times d}
\end{equation}
\begin{equation}
\mathbf{\hat{G}}_m = \mathbf{T}_m(\mathbf{G}_m; \mathbf{\delta}_m)
\end{equation}
\begin{equation}
\mathbf{\hat{G}}_m = [\mathbf{\hat{g}}_{m,0}, \cdots, \mathbf{\hat{g}}_{m,i}, \cdots \mathbf{\hat{g}}_{m,N}] \in \mathbb{R}^{N\times d}
\end{equation}
where $\mathbf{\delta}_m$ denotes the parameters of the Transformer $\mathbb{T}_m(\cdot)$ for modality $m$. Therefore, we can obtain ${\mathbf{\hat{g}}_{m,i}}$ to represent the enhanced shared feature of $i$-th utterance in modality $m$.

Finally, we obtain the refined feature representation $\mathbf{\hat{x}}_{m,i}\in \mathbb{R}^{2d}$ of utterance $i$ in modality $m$, as follows: 
\begin{equation}
    \mathbf{\hat{x}}_{m,i} =[\mathbf{\hat{g}}_{m,i}, \mathbf{\hat{h}}_{m,i}].
\end{equation}
\subsection{Emotion Recognition}
Here, we integrate the final complete features ($\mathbf{\hat{x}}_{a,i}, \mathbf{\hat{x}}_{t,i}, \mathbf{\hat{x}}_{v,i}$) of all three modalities (audio, text and visual), as the input of a classifier $\mathbf{C}(\cdot)$ to give the predict result of the emotion $\hat{y}_{i}$, as follows:
\begin{equation}
    \hat{y}_{i} = \mathbf{C}([\mathbf{\hat{x}}_{a,i},\mathbf{\hat{x}}_{t,i},\mathbf{\hat{x}}_{v,i}])
\end{equation}
where the classifier $\mathbf{C}(\cdot)$ consists of a fully-connected layer and a subsequent 2-layer MLP with a ReLU, as in \cite{shi2023multiemo}.

For model optimization, our AO-FL is trained through a joint objective function $\mathcal L_{Total}$,  which consists of a cross-entropy loss ${\mathcal{L}}_{C}$, and our adaptive angle optimization strategy, as follows,
\begin{equation}
\mathcal L_{Total} =  \underbrace{\alpha \cdot {\mathcal{L}}_{CEN} + \beta \cdot {\mathcal{L}}_{ARE}}_{AAO\;\; Strategy} + \eta \cdot {\mathcal{L}}_{C}(\hat{y}_{i}, y_i)
\end{equation}
where $y_i$ is the groundtruth emotion label corresponding to $\hat{y}_i$. $\alpha$, $\beta$ and $\eta$ are hyper-parameters. To ensure stable predictions of $\theta_{m, i}$, we adopt a warm-up strategy \cite{he2016deep}, where only $\mathcal{L}_{C}$ is utilized during warm-up phase.

As a final note, our AO-FL serves as a multimodal feature disentanglement and enhancement method that can be seamlessly integrated with various unimodal feature extractors for different tasks (i.e., MER and MERC), offering both strong performance and good generalization, as discussed in Section \ref{sec:MER}.

\begin{table}[b]
\caption{Emotion label distribution on both IEMOCAP and MELD datasets.}
\label{count}
\centering
\scalebox{0.95}{
\begin{tblr}{
  rowsep = 1pt,
  colsep =1.4pt,
  cells = {c},
  cell{1}{1} = {r=2}{},
  cell{3}{1} = {r=2}{},
  vline{2} = {1,2,3,4}{},
  hline{1,3,5} = {-}{},
  hline{2,4} = {2-8}{},
}
\hline
IEMOCAP & Happy   & Sad      & Neutral & Angry   & Excited & Frustrated & -     \\
        & 648     & 1084     & 1708    & 1103    & 1041    & 1849       & -     \\
        \hline
MELD    & Neutral & Surprise & Fear    & Sadness & Joy     & Disgust    & Anger \\
        & 6202    & 1561     & 347     & 944     & 2228    & 334        & 1513  \\
\hline    
\end{tblr}}
\end{table}

\section{Experiments}
\label{sec:experiment}
\subsection{Experimental Setup}
\noindent\textbf{Dataset:} We conduct our experiments on two widely-used MERC benchmarks, i.e., IEMOCAP \cite{busso2008iemocap} and MELD \cite{poria-etal-2019-meld}. IEMOCAP \cite{busso2008iemocap} has dyadic conversation videos with ten speakers, consisting of $7,433$ utterances across $151$ dialogues in $5$ sessions, labeled with $6$ emotions: Happy, Sad, Angry, Excited, Frustrated, and Neutral. The training set uses the first four sessions, the last session for testing, and $10\%$ random training dialogues as validation. MELD \cite{poria-etal-2019-meld} includes over $1,400$ dialogues and $13,000$ utterances from `Friends' TV series. Each utterance is categorized into seven emotions: Neutral, Surprise, Fear, Sadness, Joy, Disgust, and Anger. We adopt the predefined dataset splits for training, validation, and testing. The emotion distributions of both datasets are summarized in Table \ref{count}.

\noindent\textbf{Baselines: }
To evaluate the effectiveness of AO-FL, we compare it with representative state-of-the-art methods.  Specifically, we consider widely-used non-disentangled methods, including  DialogueRNN\cite{majumder2019dialoguernn}, graph-based methods (i.e., DialogueGCN \cite{ghosal-etal-2019-dialoguegcn}, MMGCN \cite{wei2019mmgcn}, MMDFN \cite{hu2022mm} and MGLRA\cite{meng2024masked}) and Transformer-based methods (i.e., CTNet\cite{lian2021ctnet}, SCMM\cite{yang2023self}, MultiEMO\cite{shi2023multiemo} and SACCMA\cite{guo2024speaker}).  These non-disentangled methods aggregate features from all modalities based on semantic similarity. For disentanglement methods, we compare it with Joyful \cite{li-etal-2023-joyful}, which aggregates only shared features from all modalities while discarding modality-specific ones, and D$^2$GNN \cite{dai2024multimodal}, which separates emotion-relevant from emotion-irrelevant patterns within unimodal features.
\noindent\textbf{Evaluation Metrics: }
We adopt two widely used evaluation metrics for performance evaluation, including Accuracy (Acc) \cite{yang2023self,guo2024speaker} and Weighted F1-score (w-F1)
\cite{hu2022mm,li-etal-2023-joyful}, following the state-of-the-art methods \cite{meng2024masked,guo2024speaker}.

\begin{table*}
\centering

\caption{Performance comparison on IEMOCAP and MELD test dataset under the multimodal setting (text, audio and vision). \textbf{Bold} is the best, \uline{underlined} is the second best. $^{\flat }$ denotes reproduced results.}
\label{baseline}
\scalebox{1}{
\begin{tblr}{
  width = \linewidth,
  rowsep = 1pt,
  colsep = 1.6pt,
  cells = {c},
  cell{1}{1} = {r=2}{},
  cell{1}{2} = {c=8}{0.399\linewidth},
  cell{1}{10} = {c=9}{0.447\linewidth},
  vline{3} = {1}{},
  vline{8,10,17} = {2-14}{},
  hline{1} = {0.4pt}, 
    hline{3,14} = {0.4pt}, 
    hline{16} = {0.4pt}, 
  hline{2} = {2-18}{},
}
\hline
Methods      & IEMOCAP &       &         &       &         &            &       &       & MELD    &          &       &         &       &         &       &       &       \\
            & Happy   & Sad   & Neutral & Angry & Excited & Frustrated & Acc   & w-F1  & Neutral & Surprise & Fear  & Sadness & Joy   & Disgust & Anger & Acc   & w-F1  \\
DialogueRNN \cite{majumder2019dialoguernn}  & 32.20    & 80.26 & 57.89   & 62.82 & 73.87   & 59.76      & 63.52 & 62.89 & 76.97   & 47.69    &   -    & 20.41   & 50.92 &  -       & 45.52 & 60.31 & 57.66 \\
DialogueGCN\cite{ghosal-etal-2019-dialoguegcn}  & 51.57   & 80.48 & 57.69   & 53.95 & 72.81   & 57.33      & 63.22 & 62.89 & 75.95   & 46.05    & -      & 19.60    & 51.20  &  -       & 40.83 & 58.62 & 56.36 \\
CTNet \cite{lian2021ctnet}   & 51.30    & 79.90  & 65.80    & 67.20  & \uline{78.70}    & 58.80       & 68.00    & 67.50  & 77.40    & 52.70     &10.00   & 32.50    & 56.00    & 11.20    & 44.60  & 62.00    & 60.50  \\
MMGCN  \cite{wei2019mmgcn}   & 42.34   & 78.67 & 61.73   & 69.00    & 74.33   & 62.32      & 66.36 & 66.22 & 76.33   & 48.15    &   -    & 26.74   & 53.02 &  -       & 46.09 & 60.42 & 58.31 \\
MMDFN  \cite{hu2022mm}   & 42.22   & 78.98 & 66.42   & \uline{69.77} & 75.56   & 66.33      & 68.21 & 68.18 & 77.76   & 50.69    & -     & 22.93   & 54.78 & -       & 47.82 & 62.49 & 59.46 \\
SCMM \cite{yang2023self}     & 
45.37   & 78.76 & 63.54   & 66.05 & 76.70   & 66.18      & -      & 67.53 &    -     &     -     &    -   &   -      &    -   &  -       &  -     &  -     & 59.44 \\
Joyful  \cite{li-etal-2023-joyful}   & 60.94   & \uline{84.42} & 68.24   & \textbf{69.95} & 73.54   & 67.55      & 70.55 & 71.03 & 76.80    & 51.91    & -     & 41.78   & 56.89 & -       & 50.71 & 62.53 & 61.77 \\
MultiEMO  \cite{shi2023multiemo}$^{\flat }$    & 56.60   & 82.73 & 68.63  & 68.26 & 75.43   & \textbf{69.90}      & 71.10 & \uline{71.20} & 79.68    & 57.53    & \uline{20.25}    & \textbf{43.24}   & 64.06 & \uline{26.09}       & \textbf{55.06} & \uline{67.09} & \uline{66.20} \\
SACCMA \cite{guo2024speaker}     & 38.60    & \textbf{86.53} & 64.90    & 64.56 & 74.52   & 62.99      &  -     & 67.10  &  -       &   -       &  -     &   -      &  -     &  -       &     -  &   -    & 59.30  \\
D$^{2}$GNN \cite{dai2024multimodal}     & 61.11    & 83.19 & 68.22    & 66.12  & 75.22   & 63.73      &  70.22     & 69.77  &  76.38       &   49.91     &  -    &   32.18      &  56.86     &  -     &   47.60    &   61.72   & 59.74 \\
MGLRA \cite{meng2024masked}     & \uline{63.50}   & 81.50  & \uline{71.50}    & 61.10  & 76.30    & 67.80      & \uline{71.30}  & 70.10  & \textbf{80.80}    & \textbf{59.50}     & -     & 27.80    & \textbf{66.50}  & -       & 48.04 & 66.40  & 64.90  \\
\textbf{AO-FL(Ours)}        &  \textbf{64.69}  & 82.25 & \textbf{72.99}   & 67.59 & \textbf{78.77}  & \uline{68.32}     &  \textbf{72.77} & \textbf{73.05}  & \uline{79.74}   & \uline{59.35}     & \textbf{21.05} & \uline{43.15}  & \uline{65.01} & \textbf{26.92}   & \uline{52.87} & \textbf{67.47} & \textbf{66.31} \\
\hline
\end{tblr}}
\end{table*}

\subsection{Performance Comparison}
We compare AO-FL with several state-of-the-art methods on the IEMOCAP and MELD datasets, and results are given in Table \ref{baseline}. Our AO-FL achieves the best performance in terms of both Acc and w-F1 on both datasets, particularly in recognizing the Happy, Neutral and Excited emotions on the IEMOCAP dataset, and the Fear and Disgust emotions on the MELD dataset. 

Compared with non-disentangled methods (e.g., DialogueGCN \cite{ghosal-etal-2019-dialoguegcn}, MMGCN \cite{wei2019mmgcn}, and MMDFN \cite{hu2022mm}), which aggregate multimodal features based solely on semantic similarity, AO-FL significantly outperforms them across all evaluation metrics.
Moreover, in comparison to disentangled approaches such as Joyful \cite{li-etal-2023-joyful} and D$^{2}$GNN \cite{dai2024multimodal}, our AO-FL achieves superior overall recognition performance.
For instance, on both IEMOCAP and MELD datasets, AO-FL consistently yields better recognition performance across nearly all emotions.

These results further highlight the effectiveness of AO-FL in capturing both shared and specific features, while preserving their complementary semantics through partial disentanglement based on their optimal angular relationship.

\subsection{Ablation Study}
We conduct ablation studies to evaluate the effectiveness of the Adaptive Angle Optimization (AAO) strategy and the Orthogonal Projection Refinement (OPR) module, by incrementally removing these components from the AO-FL framework. The results are presented in Table \ref{ablation_AAO_OPR}.

\begin{table}[htpb!]
\caption{Ablation study of AO-FL using adaptive angle optimization (AAO) strategy and orthogonal projection refinement (OPR) module.}.
\label{ablation_AAO_OPR}
\centering
\scalebox{0.9}{
\begin{tblr}{
  width = \linewidth,
  rowsep = 1pt,
  colsep = 6.8pt,
  colspec = {Q[158]Q[152]Q[144]Q[144]Q[144]Q[144]},
  cells = {c},
  cell{1}{1} = {c=2}{0.31\linewidth},
  cell{1}{3} = {c=2}{0.288\linewidth},
  cell{1}{5} = {c=2}{0.288\linewidth},
  vline{2} = {1}{},
  vline{3} = {2-6}{},
  hline{1-3,7} = {-}{},
  hline{1,7} = {0.4pt},
}
\hline
Components &     & IEMOCAP &       & MELD  &       \\
AAO & OPR & Acc     & w-F1  & Acc   & w-F1  \\
     -      & -    & 67.96  & 68.30 & 65.67 & 64.65 \\
 $\checkmark$      &  -    & 69.13   & 69.15 & 66.51  & 65.50 \\
       -    &  $\checkmark$   & 71.41  & 71.68 & 66.28  & 65.82  \\
$\checkmark$           & $\checkmark$     & \textbf{72.77}   & \textbf{73.05} & \textbf{67.47}  & \textbf{66.31}  \\
\hline
\end{tblr}}
\end{table}

It can be observed that incorporating AAO alone leads to improvements of 1.17\% in accuracy and 0.85\% in w-F1 on the IEMOCAP dataset, and 0.84\% and 0.85\%, respectively, on the MELD dataset. These results indicate that the AAO strategy effectively enhances both shared and specific features by partially disentangling them based on their optimal angular relationship. Incorporating only the OPR module yields even greater improvements: 3.45\% in accuracy and 3.38\% in w-F1 on IEMOCAP, and 0.61\% and 1.17\% on MELD. These gains suggest that the OPR module significantly boosts emotion recognition by reducing redundancy between shared and specific features while enriching shared representations with contextual semantics.

When both AAO and OPR are combined, AO-FL achieves the best overall performance across both datasets. These findings demonstrate the effectiveness of both components, which provide complementary benefits to our framework.

\begin{figure*}
\subfloat[AO-FL (w/o -AAO)]{\label{tsne_before}\includegraphics[width=4.2cm]{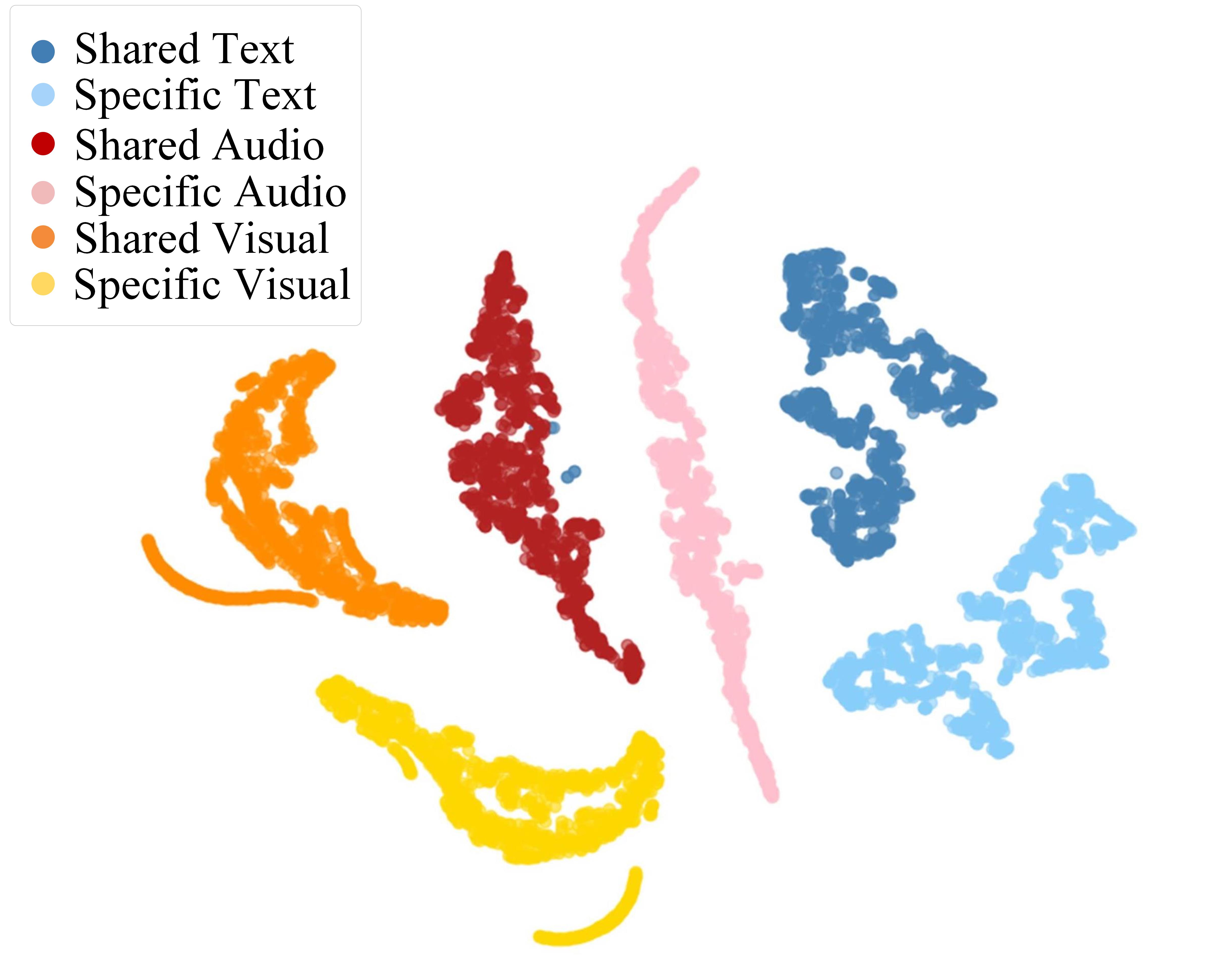}}
\subfloat[AO-FL (w/o -CEN)]{\label{tsne_after}\includegraphics[width=4.2cm]{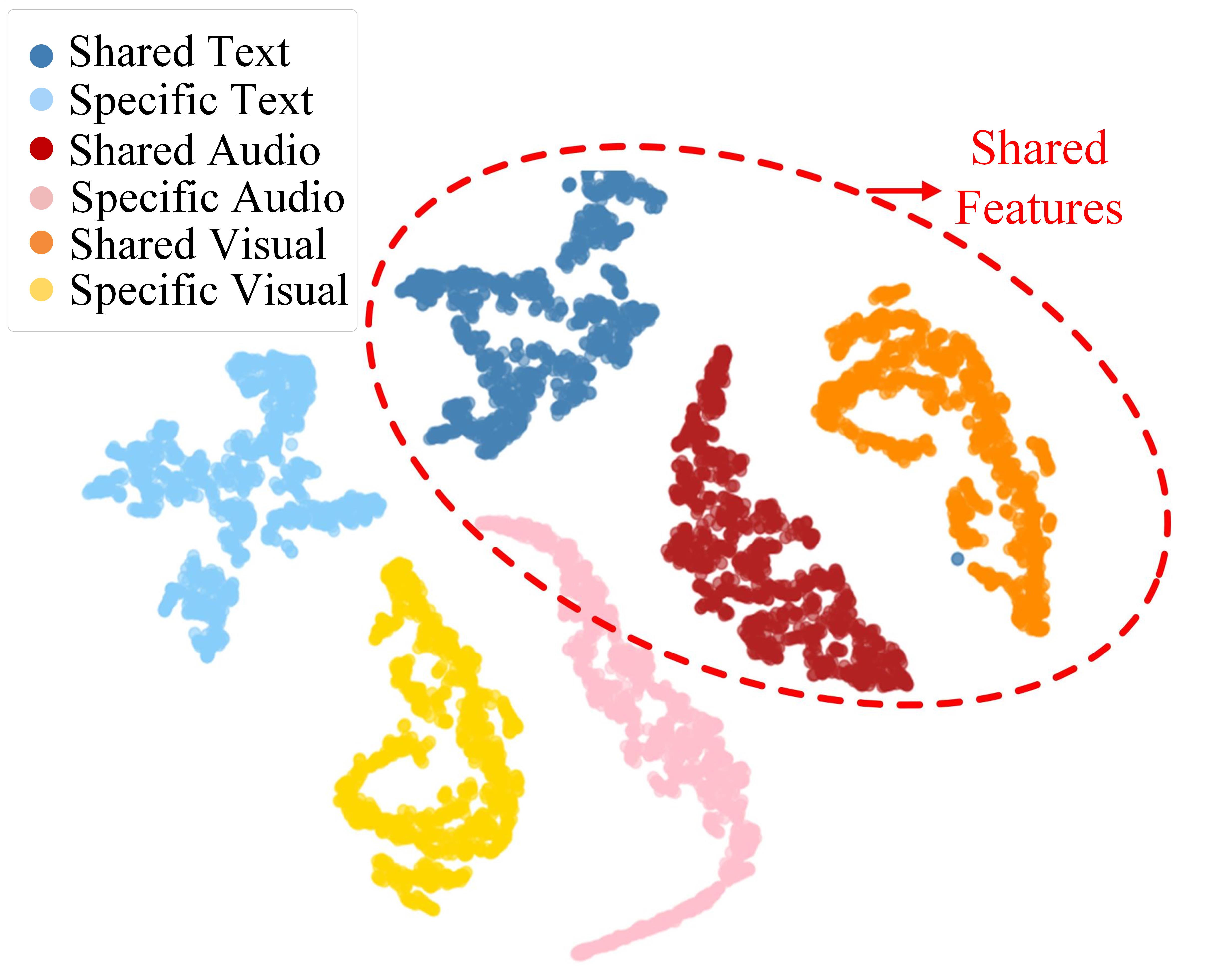}}
\subfloat[AO-FL (w/o -ARE)]{\label{tsne_after}\includegraphics[width=4.2cm]{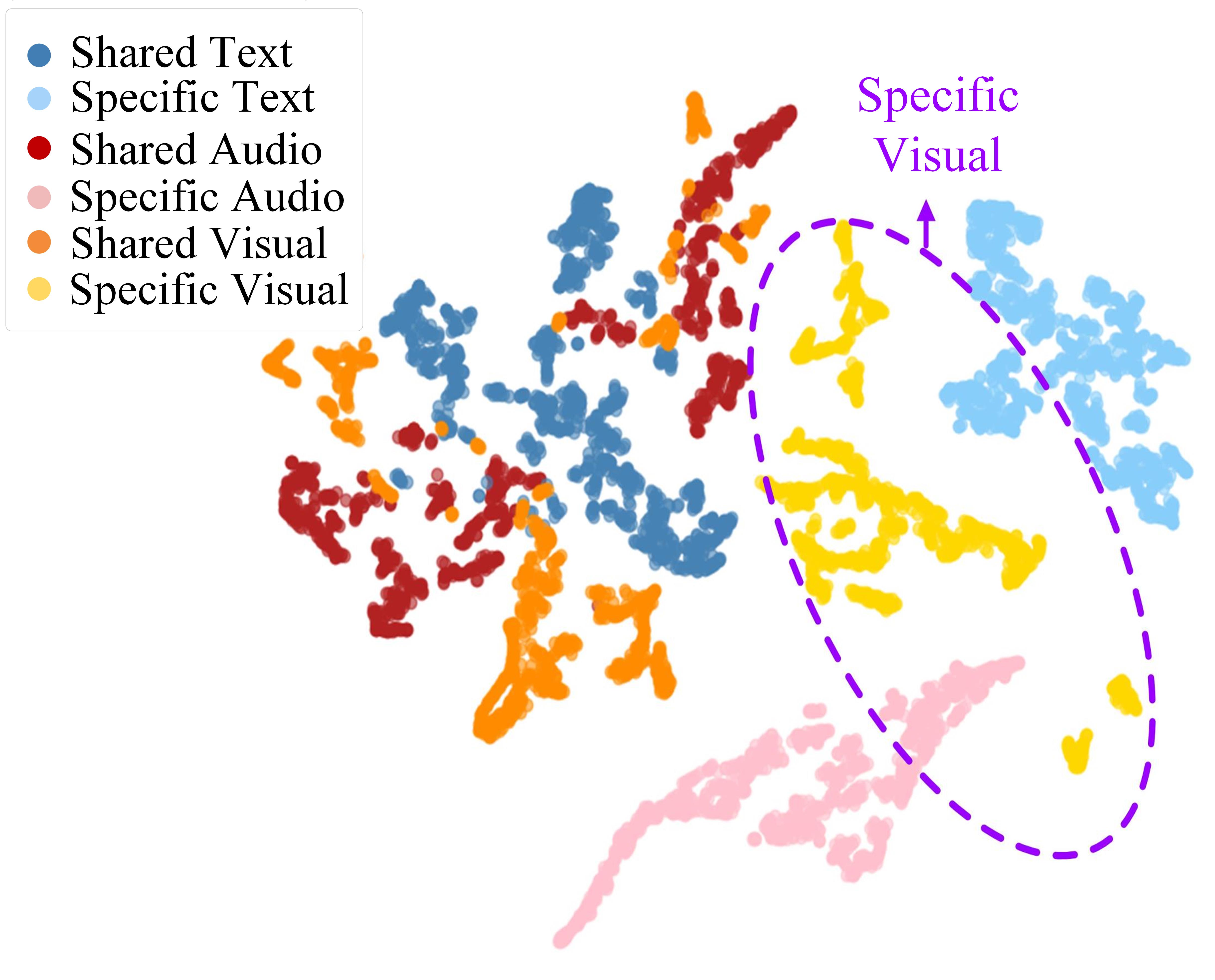}}
\subfloat[AO-FL]{\label{tsne_after}\includegraphics[width=4.2cm]{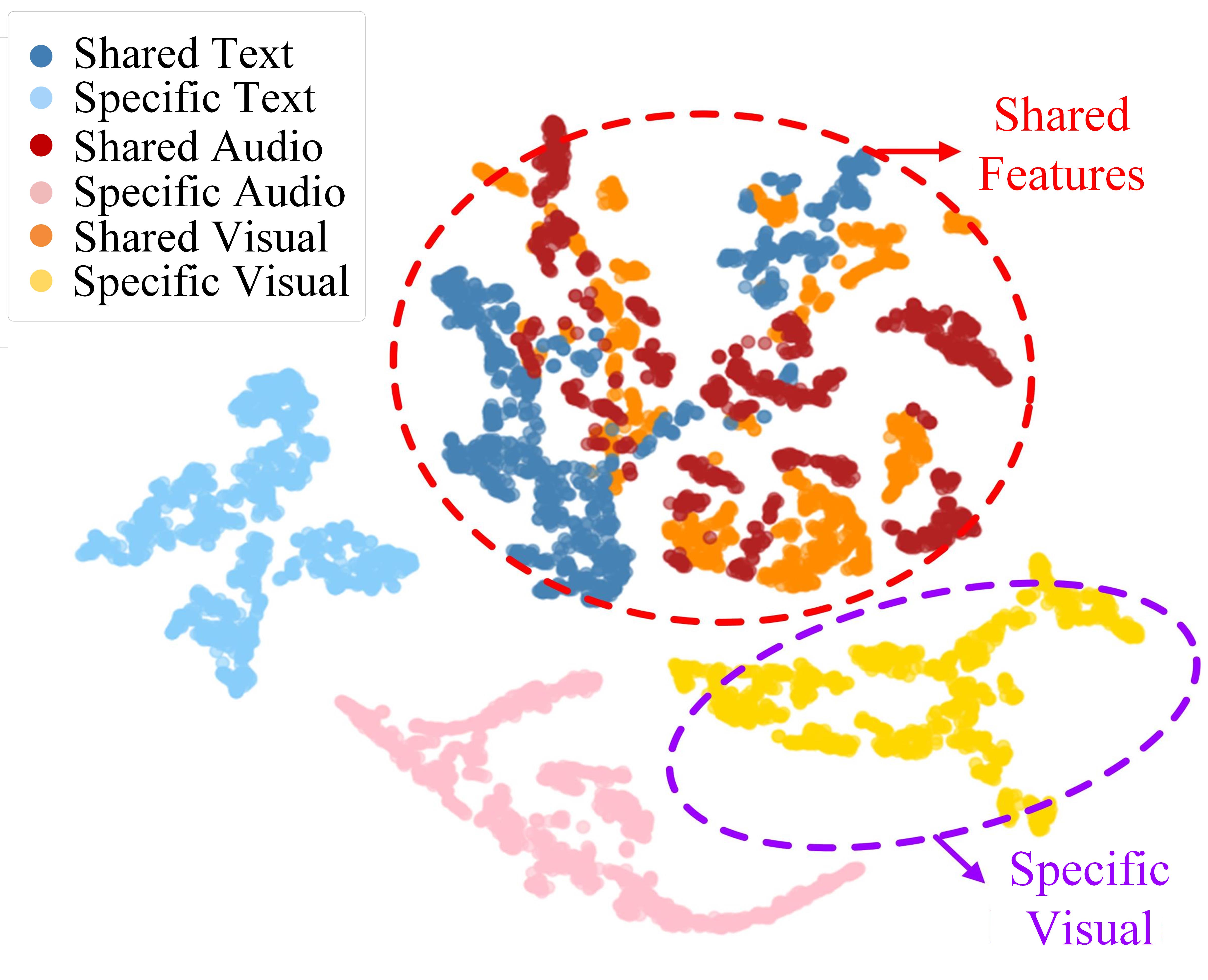}}
\caption{Visualizations of partially disentangled shared and specific features of text, audio and visual modalities on IEMOCAP dataset obtained by our AO-FL and three variants of our AO-FL, 
i.e., (a) AO-FL (w/o -AAO), (b) AO-FL (w/o -CEN), (c) AO-FL (w/o -ARE).}
\label{fig:EX_IM}
\end{figure*}

\subsection{Validation of Partial Disentanglement}
To further assess the effectiveness of our Angular Relationship Exploration (ARE) module in partially disentangling shared and specific features, we replace it with two rigid orthogonal constraints: a squared Frobenius norm constraint from MISA \cite{hazarika2020misa} and a cosine similarity constraint from DMD \cite{li2023decoupled}, both enforcing a fixed $90^\circ$ angle between two feature types to achieve full disentanglement. These variants are denoted as AO-FL (-OrtNorm) and AO-FL (-OrtCos), respectively, with the rest of the AO-FL architecture unchanged. Table~\ref{tab:Partial_D} presents a comparison between our partial disentanglement AO-FL and these full disentanglement variants.

\begin{table}[htbp!]
\centering
 \caption{Validation of partial disentanglement between shared and specific features in our AO-FL on both datasets.}
 \label{tab:Partial_D}
  \scalebox{0.9}{
    \tabcolsep=11.5pt
    
\renewcommand\arraystretch{1.2}
    {\begin{tabular}
    { l| c c|c c}
    \hline
    \hline
    \multirow{2}*{Model} & \multicolumn{2}{c|}{IEMOCAP}  & \multicolumn{2}{c}{MELD} \\
    \cline{2-5}
    &Acc & w-F1  &Acc & w-F1\\
   \cline{1-5}
    AO-FL (-OrtNorm)     & 70.67  & 70.99
                            & 66.28  & 65.52 \\
    AO-FL (-OrtCos)      & 69.32  & 69.39
                         & 66.59  & 65.40 \\
    \hline
    \textbf{AO-FL}     & \textbf{72.77}   & \textbf{73.05} 
                      & \textbf{67.47}  & \textbf{66.31} \\

    \hline
    \hline
\end{tabular}}}
\end{table}

As shown in Table \ref{tab:Partial_D}, AO-FL achieves the highest accuracy and w-F1 on both IEMOCAP and MELD. 
AO-FL outperforms both AO-FL (-OrtNorm) and AO-FL (-OrtCos), 
which enforce a fixed orthogonal ($90^\circ$) relationship between these two types of feature. This illustrates that such a rigid orthogonal relationship overemphasizes distinctiveness, and neglects complementary emotion-relevant cues.
In contrast, our ARE module learns adaptive angular relationships ($0^\circ$–$180^\circ$), enabling flexible disentanglement that balances distinctiveness and complementarity, thus proving the effectiveness of AO-FL in MERC.

\subsection{Analysis of AAO Strategy}
We analyze the effectiveness of the AAO strategy by individually ablating its two components: Consistency Enhancement (CEN) function and Angular Relationship Exploration (ARE). Here, the components of our AAO strategy are indicated in abbreviated form, i.e., `-CEN' indicates the CEN function, and `-ARE’ refers to the ARE function. In addition, `w/o' is applied to indicate without specified components. The results are given in Table \ref{tab:ablation_AAO}.
\begin{table}[htbp!]
\centering
 \caption{Validation of the consistency enhancement function and the angular relationship exploration (ARE) module within adaptive angle optimization (AAO) strategy on both datasets.}
  \label{tab:ablation_AAO}
  \scalebox{0.9}{
    \tabcolsep=11.5pt
\renewcommand\arraystretch{1.2}
    {\begin{tabular}
    { l| c c|c c}
    \hline
    \hline
    \multirow{2}*{Model} & \multicolumn{2}{c|}{IEMOCAP}  & \multicolumn{2}{c}{MELD} \\
    \cline{2-5}

    &Acc & w-F1  &Acc & w-F1\\
   \cline{1-5}
    AO-FL (w/o -CEN)      & 71.72  & 71.86
                         & 66.59  & 66.28 \\
    AO-FL (w/o -ARE)     & 72.09   & 72.37
                        & 66.48  & 66.15\\
    \cline{1-5}
    \textbf{AO-FL}     & \textbf{72.77}   & \textbf{73.05} 
                      & \textbf{67.47}  & \textbf{66.31} \\

    \hline
    \hline
\end{tabular}}}
\label{tab:ablation_AAO}
\end{table}

As shown in Table \ref{tab:ablation_AAO}, AO-FL outperforms AO-FL (w/o -CEN), confirming the importance of CEN in aligning shared features across modalities to capture consistent cross-modal semantics for MERC. Similarly, AO-FL surpasses AO-FL (w/o -ARE) on all metrics, demonstrating the effectiveness of ARE in modeling the angular relationship between shared and specific features with an adaptive partial disentanglement angle to preserve both distinctiveness and complementary cues.  
The overall best performance of AO-FL validates the effectiveness of the AAO strategy. A visual analysis is provided in the next section.
\subsection{Visualization of AAO Strategy}

To further assess the effectiveness of the AAO strategy in partially disentangling shared and specific features,  we present t-distributed Stochastic Neighbor Embedding (t-SNE) visualizations \cite{van2008visualizing} in Fig.~\ref{fig:EX_IM}.
These visualizations illustrate the distribution of disentangled features from the textual, audio, and visual modalities on the IEMOCAP dataset. 
Fig.~\ref{fig:EX_IM}(d) presents the results of our full AO-FL model, while in Fig.~\ref{fig:EX_IM} (a–c), we show results from three ablated variants. The abbreviations `-CEN', `-ARE', and `-AAO' denote the consistency enhancement function, the ARE function, and the entire AAO strategy, respectively. `w/o' denotes AO-FL  without specified components.

\begin{figure*}[ht]
\subfloat[Performance with changing $\alpha$]{\label{tsne_after}\includegraphics[width=4.45cm]{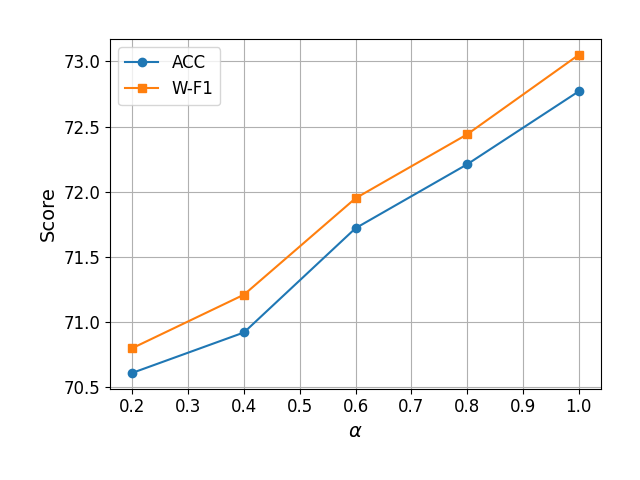}}
\subfloat[Performance with changing $\beta$]{\label{tsne_after}\includegraphics[width=4.45cm]{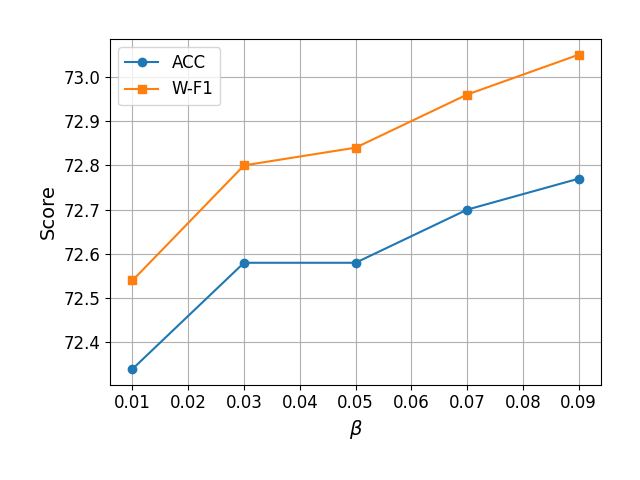}}
\subfloat[Performance with changing $\gamma$ ]{\label{tsne_after}\includegraphics[width=4.45cm]{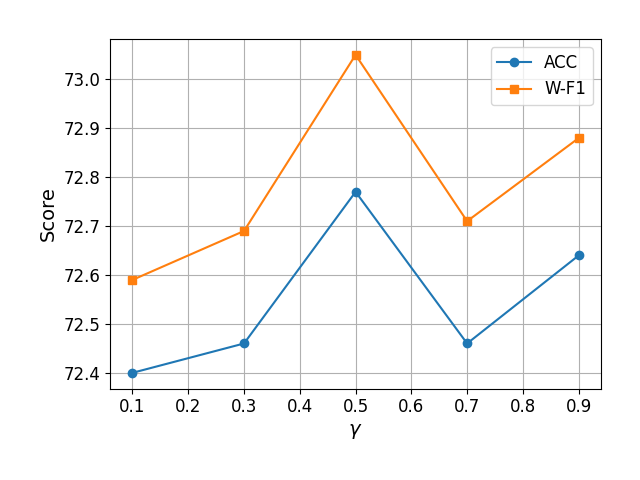}}
\subfloat[Performance with changing $\mu$]{\label{tsne_before}\includegraphics[width=4.45cm]{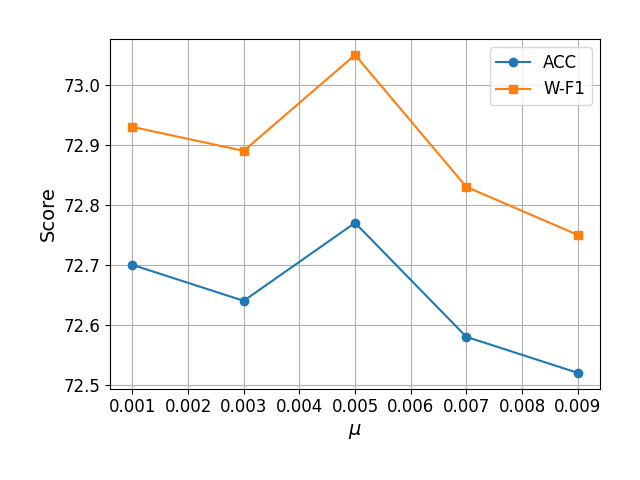}}
\caption{The performance of our AO-FL on the IEMOCAP dataset with changing hyperparameters (i.e, $\mu$, $\beta$, $\alpha$, and $\gamma$).
}
\label{para}
\end{figure*}

As shown in Fig.~\ref{fig:EX_IM}, the shared features from all three modalities in (d) AO-FL form a more unified and compact cluster compared to those in (a) AO-FL (w/o -AAO) and (b) AO-FL (w/o -CEN). This indicates that CEN function effectively aligns shared features across modalities, facilitating the capture of cross-modal consistent semantics.

In addition, the visual specific features in (d) AO-FL are more compactly clustered than those in (c) AO-FL (w/o -ARE), indicating that ARE effectively disentangles specific features to capture modality-specific semantics. However, they are less compact than those in (a) AO-FL (w/o -AAO), suggesting that ARE retains emotion-related information with shared features as complementary information. In contrast, specific features in AO-FL (w/o -AAO) are more distinctly separated from other features due to the lack of shared-specific constraints.

These observations further confirm that implicit features in AO-FL are partially disentangled from their corresponding shared features, maintaining a balance between modality-specific distinctiveness and cross-modal consistency. Overall, these findings strongly demonstrate that AO-FL, equipped with the AAO strategy, effectively achieves partial disentanglement of shared and specific features, contributing to improved emotion recognition performance.

\subsection{Analysis of ARE Module}
To validate the effectiveness of our Angular Relationship Exploration (ARE) module, which consists of the Adaptive Angle Constraint (AAC) function and the Cosine Similarity Ranking (CSR) criterion, we conduct further ablation studies. For brevity, we denote these components as `-AAC' and `-CSR', respectively. `w/o' represents AO-FL without specified components. The experimental results are presented in Table~\ref{tab:ablation_ARO}.
\begin{table}[htbp!]
\centering
 \caption{Validation of adaptive angle constraint (AAC) function and cosine similarity rank (CSR) function of our angular relationship exploration (ARE) module on both datasets.}
  \label{tab:ablation_ARO}
  \scalebox{0.9}{
    \tabcolsep=11.5pt
\renewcommand\arraystretch{1.2}
    {\begin{tabular}
    { l| c c|c c}
    \hline
    \hline
    \multirow{2}*{Model} & \multicolumn{2}{c|}{IEMOCAP}  & \multicolumn{2}{c}{MELD} \\
    \cline{2-5}

    &Acc & w-F1  &Acc & w-F1\\
   \cline{1-5}
    AO-FL (w/o -AAC)      & 72.27  & 72.56
                         & 66.86  & 65.89 \\
    AO-FL (w/o -CSR)       & 72.34   & 72.61
                        & 66.93 & 65.97\\
    \cline{1-5}
    \textbf{AO-FL}     & \textbf{72.77}   & \textbf{73.05} 
                      & \textbf{67.47}  & \textbf{66.31} \\
    \hline
    \hline
\end{tabular}}}
\end{table}

It can be observed that all metrics of AO-FL (w/o -AAC) drop more significantly than AO-FL (w/o -CSR) on both datasets. This indicates the necessity of adaptively capturing the angular relationship between shared and specific features. 
Meanwhile, AO-FL achieves the best performance, validating the importance of jointly leveraging the AAC and CSR functions within our ARE module, as CSR enhances the distinctiveness between the two feature types, and avoids the learned angular relationship from becoming too small to distinguish shared and specific features.
\subsection{Application of AO-FL to Other Multimodal Tasks}
\label{sec:MER}

To further evaluate the generalization of our proposed AO-FL for multimodal feature disentanglement and enhancement, we integrate our AO-FL with advanced LLM-based unimodal feature extractors \cite{lian2024merbench} (i.e., Baichuan-13B-Base \cite{yang2023baichuan} for text, Chinese-HuBERT-Large \cite{hsu2021hubert} for audio and CLIP-ViT-Large-Patch14 \cite{radford2021learning} for vision), denoting as AO-FL (MER), to solve a broader multimodal emotion recognition (MER) task on CMU-MOSI and CMU-MOSEI datasets. 
The comparison results between our AO-FL (MER) and baselines, including MISA \cite{hazarika2020misa}, DMD \cite{li2023decoupled} and ConFEDE \cite{yang2023confede}, are given in Table~\ref{cmu}.

\begin{table}[htbp!]
\centering
 \caption{
Generalization experiments on the CMU-MOSI and CMU-MOSEI datasets for the MER task, where AO-FL is adapted with advanced LLM-based unimodal feature extractors, 
demonstrating its effectiveness and adaptability.
}
  \label{cmu}
  \scalebox{0.9}{
    \tabcolsep=11.5pt
\renewcommand\arraystretch{1.2}
    {\begin{tabular}
    { l| c c|c c}
    \hline
    \hline
    \multirow{2}*{Model} & \multicolumn{2}{c|}{CMU-MOSI}  & \multicolumn{2}{c}{CMU-MOSEI} \\
    \cline{2-5}

    &Acc & w-F1  &Acc & w-F1\\

\cline{1-5}
    MISA\cite{hazarika2020misa}   & 78.05  & 77.84  & 83.57 & 83.61 \\
    DMD\cite{li2023decoupled}   & 75.76  & 75.89  & 84.09  & 83.91 \\
    ConFEDE\cite{yang2023confede}  & 77.59  & 77.40  & 83.79  & 83.78 \\
    \hline
    \textbf{AO-FL (MER)}     & \textbf{79.88}  & \textbf{79.72} & \textbf{84.95}  & \textbf{84.82} \\
    \hline
    \hline
\end{tabular}}}
\label{tab:ablation_AAO}
\end{table}

It can be seen that AO-FL (MER) achieves the best performance on the MER task across both CMU-MOSI and CMU-MOSEI datasets, which validates the effectiveness of our AO-FL as a multimodal feature disentanglement and enhancement module beyond the MERC task. These results also demonstrate the generalizability of AO-FL to other multimodal tasks through seamless integration with various unimodal feature extractors.

\subsection{Hyperparameter Selection}

We conduct a sensitivity analysis of four key hyperparameters (i.e., $\mu$, $\gamma$, $\alpha$, and $\beta$) on the IEMOCAP dataset, with results are shown in Fig.~\ref{para}.

As shown in Fig.~\ref{para} (a), AO-FL consistently benefits from increasing $\alpha$, underscoring the importance of aligning shared features across modalities.
In Fig.~\ref{para} (b), performance steadily improves with higher $\beta$, confirming the effectiveness of the AAO strategy for the MERC task.
Fig.~\ref{para} (c) and Fig.~\ref{para} (d) show that AO-FL achieves the best results when $\gamma = 0.5$ and $\mu = 0.005$, indicating that moderate values effectively align the actual and predicted angles between shared and specific features while preventing overly small angles, thus achieving effective partial disentanglement through the joint AAC and CSR losses.

Based on the above analysis, we empirically set $\alpha = 1$, $\beta = 0.09$, $\gamma = 0.5$, and $\mu = 0.005$ in our main experiments, as this configuration achieves optimal performance.

\begin{figure*}[ht]
\subfloat[Features of Utterance 1]{\label{tsne_before}\includegraphics[width=4.45cm]{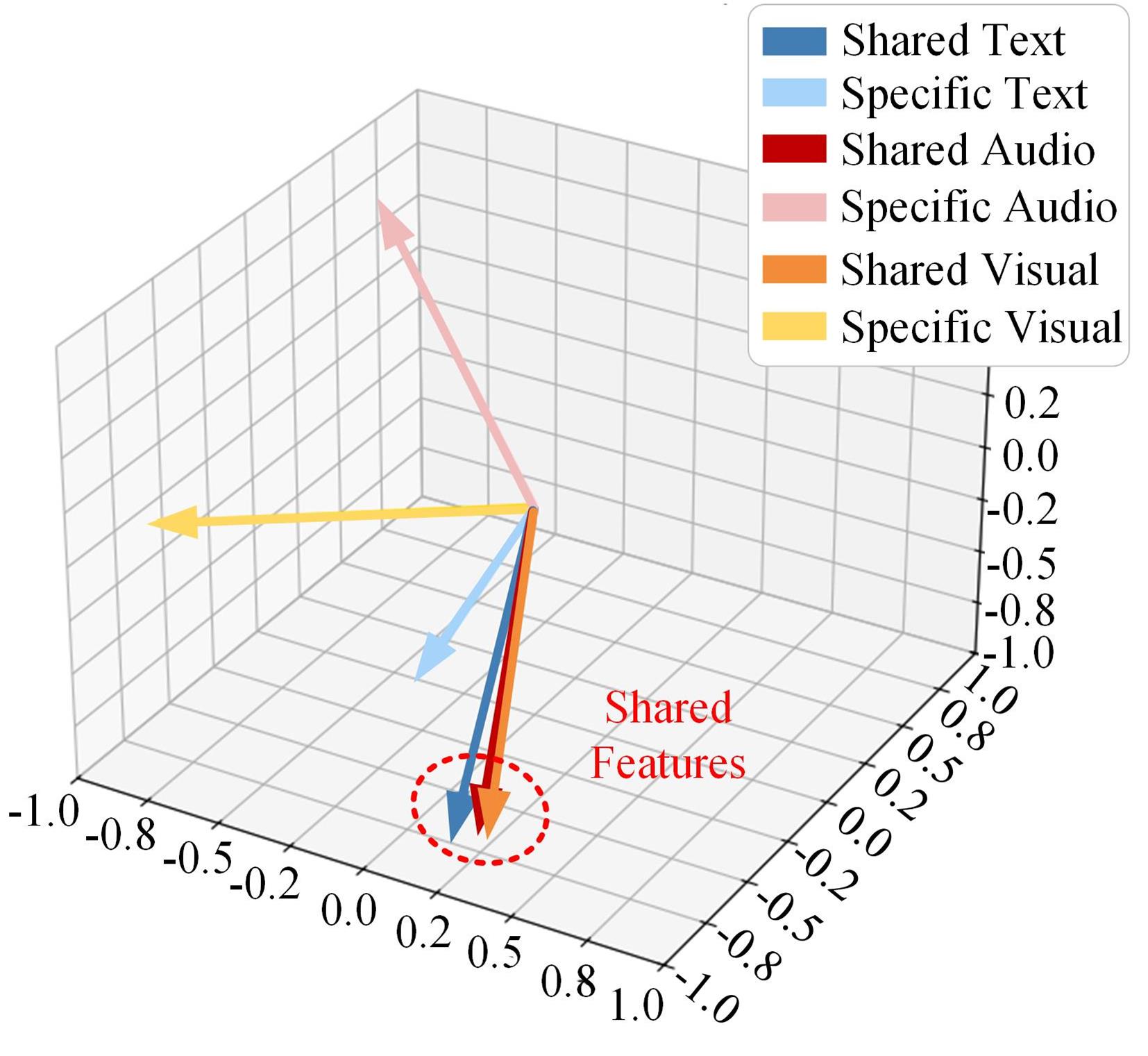}}
\subfloat[Features of Utterance 2]{\label{tsne_after}\includegraphics[width=4.45cm]{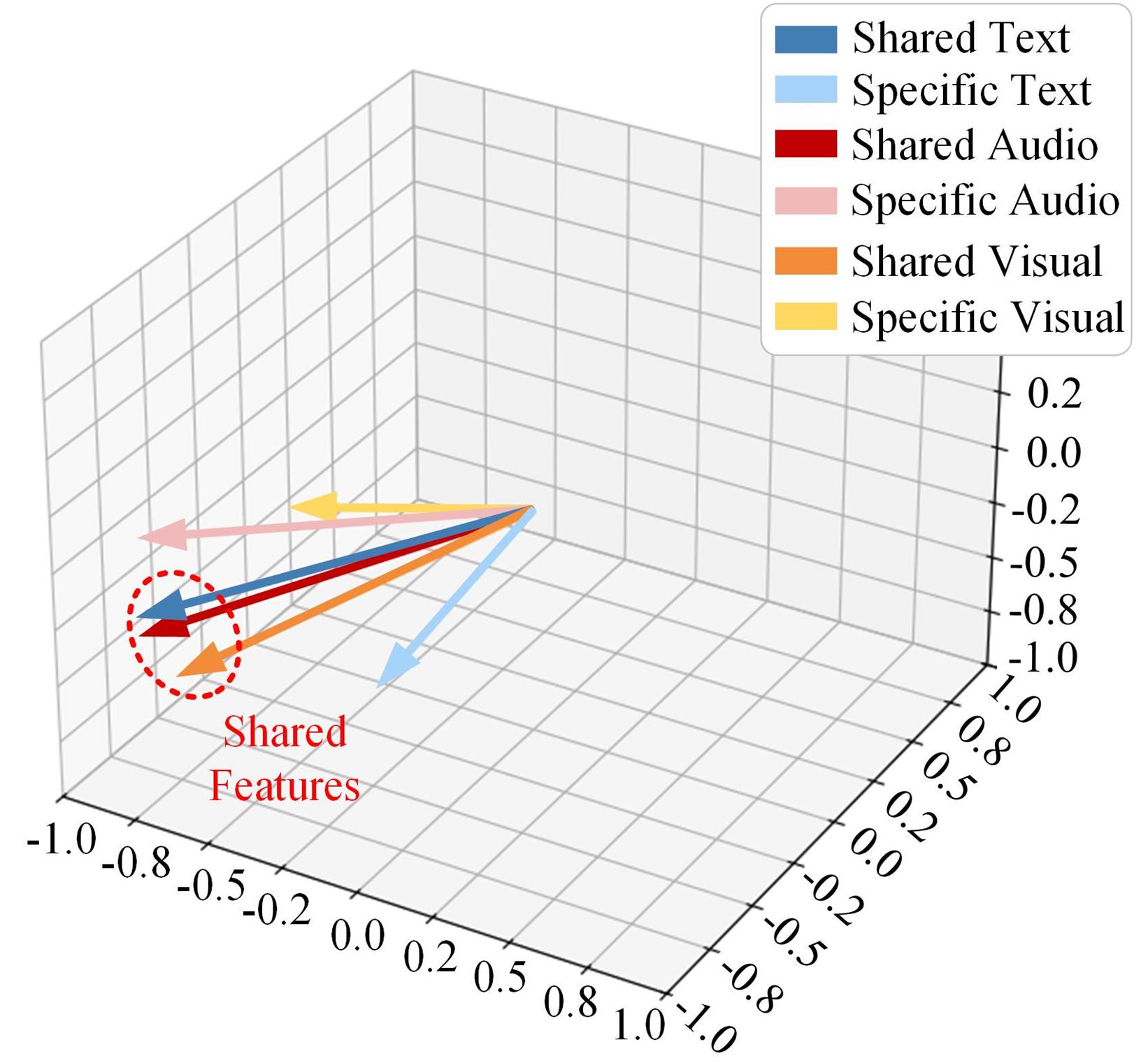}}
\subfloat[Features of Utterance 3]{\label{tsne_after}\includegraphics[width=4.45cm]{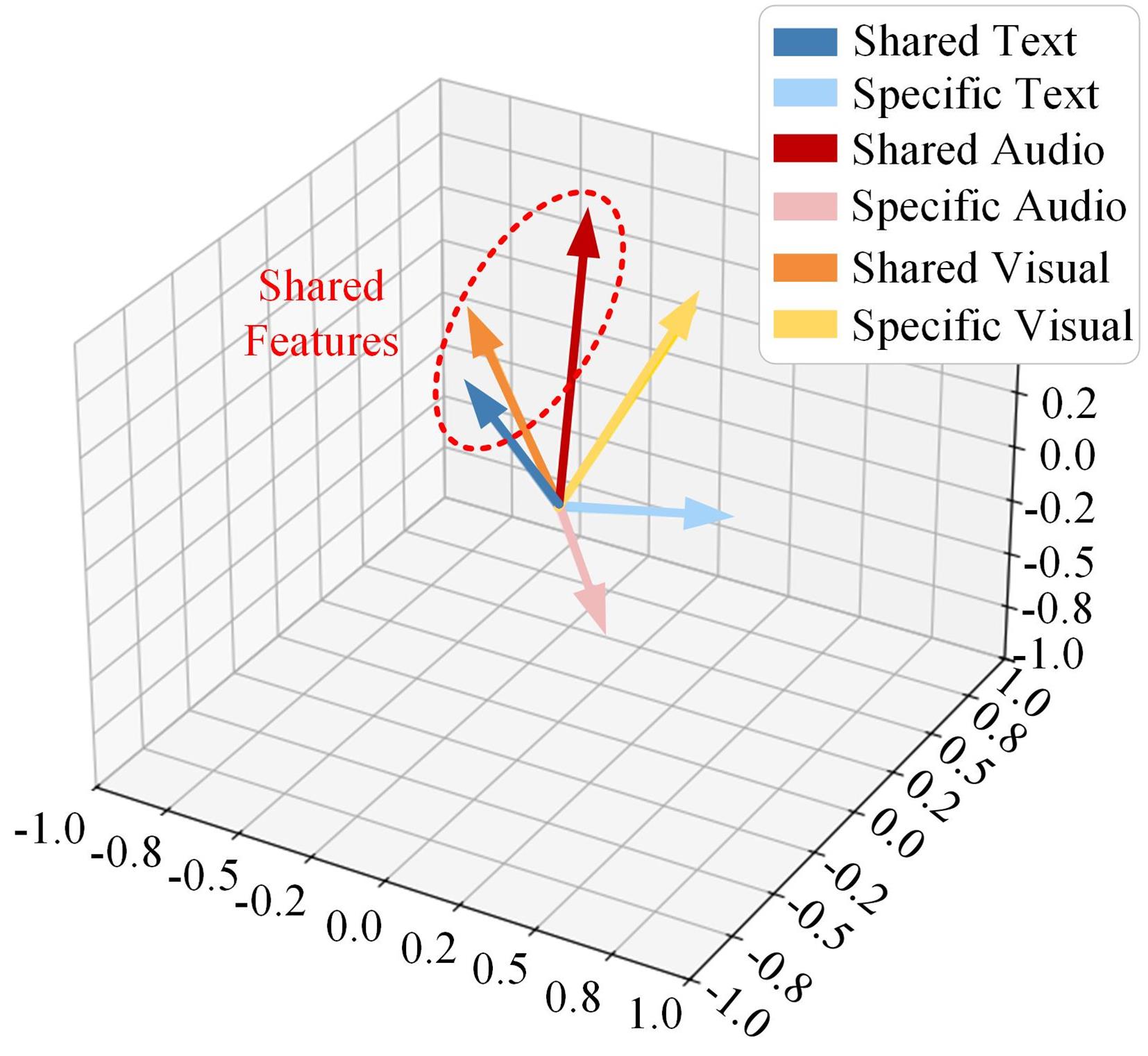}}
\subfloat[Features of Utterance 4]{\label{tsne_after}\includegraphics[width=4.45cm]{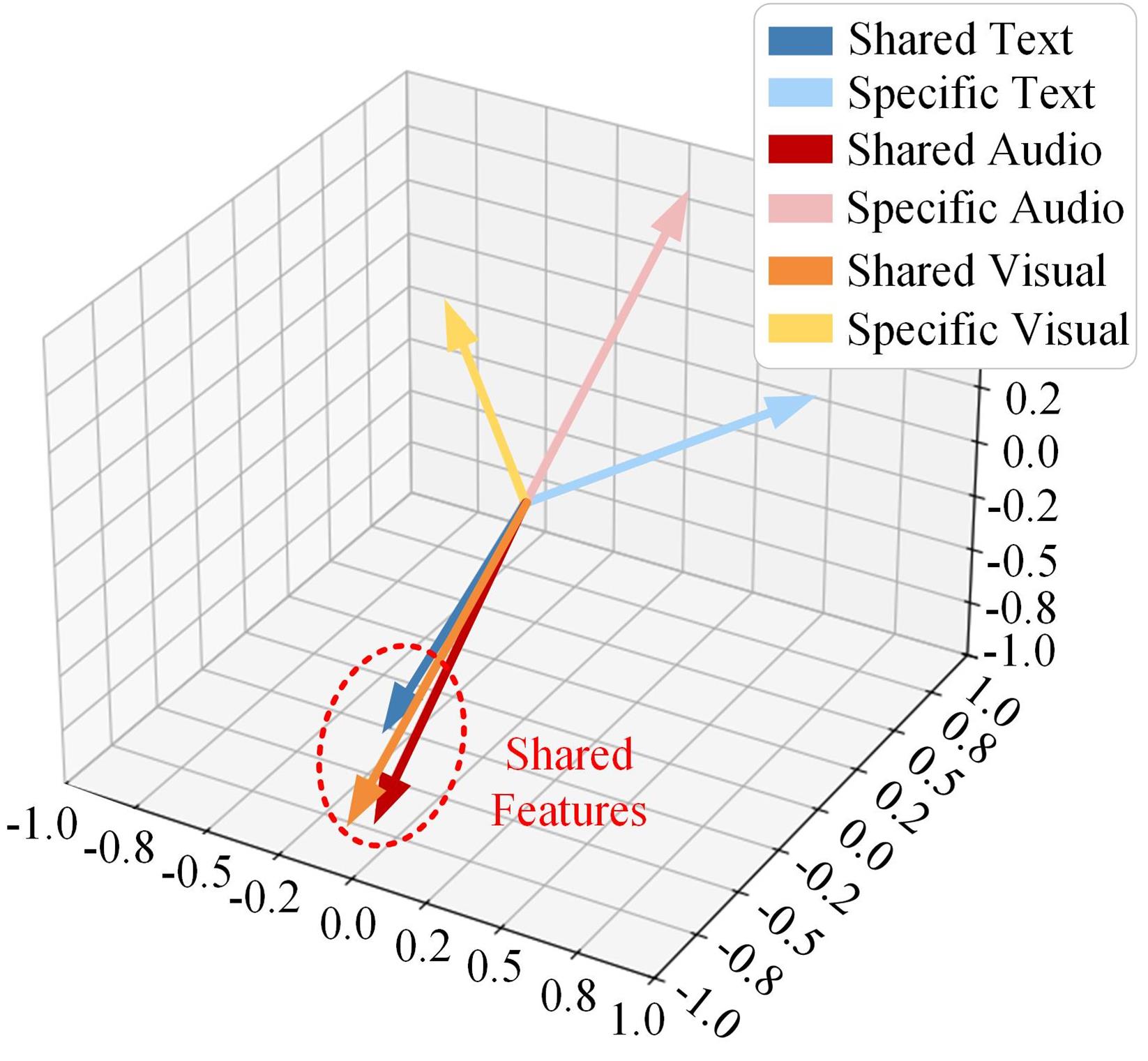}}
\caption{Partially disentangled shared and specific features of text, audio, and visual modalities for four randomly selected utterances obtained by our AO-FL model on the IEMOCAP dataset.}
\label{fig_frame}
\end{figure*}

\subsection{Disentanglement Case Study}
\label{sec:case}
To illustrate the effectiveness of our AO-FL approach in partially disentangling the shared and specific features from each modality (i.e., text, auditory, and visual), we randomly sample four different utterances from the IEMOCAP dataset and present their partially disentangled results in Fig.~\ref{fig_frame}.

As shown in Fig.~\ref{fig_frame}, the shared features from all three modalities have highly similar orientations with small angles in all subfigures (a)–(d). 
This demonstrates that the CEN function effectively pulls together shared features of the same utterance across modalities as illustrated in Fig.~\ref{CE_loss}, capturing cross-modal consistent semantics. This also confirms that our AO-FL can successfully align shared features across modalities through the AAO strategy.

Additionally, within each subgraph, the angles between shared and specific features from the same modality are consistently larger than the angles between shared features from different modalities, aligning well with the constraint imposed by our CSR function as illustrated in Fig.~\ref{CSR_loss}.

Furthermore, the angles between shared and specific features within each modality range from acute angles (e.g., Fig.~\ref{fig_frame} (b) and (c)) to obtuse angles (e.g., Fig.~\ref{fig_frame} (a) and (d)). This variation demonstrates the flexibility of our Adaptive Angle Constraint (AAC) function in adaptively learning optimal angular relationships to achieve effective partial disentanglement.

Overall, these observations strongly validate the effectiveness of our AO-FL framework, which leverages adaptive angular optimization between shared and specific features to achieve partial disentanglement, thereby enhancing multimodal emotion recognition performance.

\section{Conclusion}
\label{sec:conclusion}
In this paper, we proposed the Angle-Optimized Feature Learning (AO-FL) framework, which integrated cross-modal consistent shared features with modality-specific features for MERC.
Specifically, AO-FL employed an AAO strategy to partially disentangle these two feature types within each modality, preserving both their distinctiveness and complementary cues for emotion recognition.
Additionally, an OPR module further reduced redundancy in modality-specific features and enriched shared features with contextual information, enhancing their representational quality.
Extensive experiments on the IEMOCAP and MELD datasets demonstrated AO-FL’s effectiveness for MERC, while additional evaluations on datasets of MER task, where AO-FL was integrated with different unimodal feature extractors, confirmed its adaptability and strong generalization capability.

\normalem

\bibliographystyle{IEEEtran}
\bibliography{sample-ieee}

\end{document}